\documentclass[twocolumn]{aastex631}
\usepackage{newtxtext,newtxmath}
\usepackage{nccmath}
\usepackage{soul}

\usepackage{natbib}
\bibpunct{(}{)}{;}{a}{}{,}

\usepackage[english]{babel}
\usepackage{graphicx}

\begin{document}

\title{Mitigating polarization leakage in gas pixel detectors through hybrid machine learning and analytic event reconstruction}

\author[0000-0003-3842-4493]{Nicol\'o Cibrario}
\affiliation{Istituto Nazionale di Fisica Nucleare, Sezione di Torino, Via Pietro Giuria 1, 10125 Torino, Italy}
\affiliation{Dipartimento di Fisica, Università degli Studi di Torino, Via Pietro Giuria 1, 10125 Torino, Italy}
\author[0000-0002-6548-5622]{Michela Negro}
\affiliation{Department of Physics \& Astronomy, Louisiana State University, Baton Rouge, LA 70803, USA}
\author[0000-0002-4264-1215]{Raffaella Bonino}
\affiliation{Istituto Nazionale di Fisica Nucleare, Sezione di Torino, Via Pietro Giuria 1, 10125 Torino, Italy}
\affiliation{Dipartimento di Fisica, Università degli Studi di Torino, Via Pietro Giuria 1, 10125 Torino, Italy}
\author[0000-0002-7127-1006]{Nikita Moriakov}
\affiliation{Informatics Institute, University of Amsterdam, Science Park 900, 1098 XH Amsterdam, the Netherlands}
\author[0000-0002-9785-7726]{Luca Baldini}
\affiliation{Istituto Nazionale di Fisica Nucleare, Sezione di Pisa, Largo B. Pontecorvo 3, 56127 Pisa, Italy}
\affiliation{Dipartimento di Fisica, Università di Pisa, Largo B. Pontecorvo 3, 56127 Pisa, Italy}
\author[0000-0002-7574-1298]{Niccol\'o Di Lalla}
\affiliation{Department of Physics and Kavli Institute for Particle Astrophysics and Cosmology, Stanford University, Stanford, California 94305, USA}
\author[0000-0003-0331-3259]{Alessandro Di Marco}
\affiliation{INAF – IAPS, via Fosso del Cavaliere, 100, Rome, Italy I-00133}
\author[0000-0003-1533-0283]{Sergio Fabiani}
\affiliation{INAF – IAPS, via Fosso del Cavaliere, 100, Rome, Italy I-00133}
\author[0009-0000-2764-5085]{Andrea Frass\'a}
\affiliation{Istituto Nazionale di Fisica Nucleare, Sezione di Torino, Via Pietro Giuria 1, 10125 Torino, Italy}
\author[0000-0002-8085-2304]{Alessio Gorgi}
\affiliation{Istituto Nazionale di Fisica Nucleare, Sezione di Torino, Via Pietro Giuria 1, 10125 Torino, Italy}
\author[0000-0001-8916-4156]{Fabio La Monaca}
\affiliation{INAF – IAPS, via Fosso del Cavaliere, 100, Rome, Italy I-00133}
\affiliation{Dipartimento di Fisica, Universit\'{a} degli Studi di Roma Tor Vergata, Via della Ricerca Scientifica 1, 00133 Rome, Italy}
\author[0000-0002-0984-1856]{Luca Latronico}
\affiliation{Istituto Nazionale di Fisica Nucleare, Sezione di Torino, Via Pietro Giuria 1, 10125 Torino, Italy}
\author[0000-0002-0698-4421]{Simone Maldera}
\affiliation{Istituto Nazionale di Fisica Nucleare, Sezione di Torino, Via Pietro Giuria 1, 10125 Torino, Italy}
\author[0000-0002-0998-4953]{Alberto Manfreda}
\affiliation{Istituto Nazionale di Fisica Nucleare, Sezione di Napoli, Strada Comunale Cinthia, 80126 Napoli, Italy}
\author[0000-0003-3331-3794]{Fabio Muleri}
\affiliation{INAF – IAPS, via Fosso del Cavaliere, 100, Rome, Italy I-00133}
\author[0000-0002-5448-7577]{Nicola Omodei}
\affiliation{Department of Physics and Kavli Institute for Particle Astrophysics and Cosmology, Stanford University, Stanford, California 94305, USA}
\author[0000-0002-9774-0560]{John Rankin}
\affiliation{INAF – IAPS, via Fosso del Cavaliere, 100, Rome, Italy I-00133}
\affiliation{INAF – OAB Merate, Via E. Bianchi 46, 23807 Merate, Italy}
\author[0000-0001-5676-6214]{Carmelo Sgr\'o}
\affiliation{Istituto Nazionale di Fisica Nucleare, Sezione di Pisa, Largo B. Pontecorvo 3, 56127 Pisa, Italy}
\author[0000-0002-8665-0105]{Stefano Silvestri}
\affiliation{Istituto Nazionale di Fisica Nucleare, Sezione di Pisa, Largo B. Pontecorvo 3, 56127 Pisa, Italy}
\affiliation{Dipartimento di Fisica, Università di Pisa, Largo B. Pontecorvo 3, 56127 Pisa, Italy}
\author[0000-0002-7781-4104]{Paolo Soffitta}
\affiliation{INAF – IAPS, via Fosso del Cavaliere, 100, Rome, Italy I-00133}
\author[0000-0002-3318-9036]{Stefano Tugliani}
\affiliation{Istituto Nazionale di Fisica Nucleare, Sezione di Torino, Via Pietro Giuria 1, 10125 Torino, Italy}
\affiliation{Dipartimento di Fisica, Università degli Studi di Torino, Via Pietro Giuria 1, 10125 Torino, Italy}

\begin{abstract}
Spatially resolved polarization measurements of extended X-ray sources are expanding our understanding of the emission mechanisms and magnetic field properties involved. Such measurements have been possible in the past few years thanks to the Imaging X-ray Polarimetry Explorer (IXPE). However, the analysis of extended sources suffers a systematic effect known as polarization leakage, which artificially affects the measured polarization signal. To address this issue, we built a hybrid reconstruction algorithm, which combines machine learning and analytic techniques to improve the reconstruction of photoelectron tracks in Gas Pixel Detectors (GPDs) and to significantly mitigate polarization leakage. 

This work presents the first application of this hybrid method to experimental data, including both calibration lab measurements and IXPE observational data. We confirmed the reliable performance of the hybrid method for both cases. Additionally, we demonstrated the algorithm’s effectiveness in reducing the polarization leakage effect through the analysis of the IXPE observation of the supernova remnant G21.5-0.9. By enabling more reliable polarization measurements, this method can potentially yield deeper insights into the magnetic field structures, particle acceleration processes, and emission mechanisms at work within extended X-ray sources.
\end{abstract}

\section{Introduction}

X-ray polarimetry provides a unique and powerful tool for studying the geometry, emission mechanisms, and magnetic field structures of a wide range of astrophysical sources.  

Spatially resolved astrophysical polarization measurements in the X-ray band were made possible thanks to the introduction of the Gas Pixel Detector (GPD), a gaseous detector which exploits the photoelectric effect to measure the polarization of incoming X-rays \citep{gpd}. The GPD is onboard the Imaging X-ray Polarimetry Explorer (IXPE), a NASA mission designed to perform astrophysical X-ray polarimetry in the 2-8 keV energy range and in orbit since 2021 \citep{Soffitta_2021,IXPE_prelaunch}. The imaging capabilities of the GPD enable the spatial resolution of certain astrophysical sources, facilitating spatially resolved polarization measurements. Some of these sources exhibit an induced polarization pattern associated with intensity edges and gradients, resulting from inaccuracies in reconstructing the photons' impact point position in the detector. This systematic effect is referred to as \textit{polarization leakage} \citep{leakage}.

The reconstruction of IXPE data is currently based on an analytic method, known as \textit{moment analysis} \citep{moment_analysis}, which enables the retrieval of polarization properties of the incident radiation by reconstructing the photoelectron tracks generated through interactions of the photons with the GPD gas cell. Specifically, the polarization degree (PD) and polarization angle (PA) of the incident radiation are derived from the analysis of the emission angles $\phi_i$ of the photoelectron tracks as \citep{kislat}:
\begin{equation}
\rm PD  = \frac{1}{\mu} \sqrt{\rm q^2 + u^2}
\end{equation}

\begin{equation}
\rm PA  = \frac{1}{2} \arctan\left(\frac{u}{q}\right)
\end{equation}
\noindent
where q and u are the normalized Stokes parameters of the observation.
\begin{equation}
\rm
q = \frac{2}{N}\sum_{i=1}^N \cos 2\phi_i \-\ \-\ \-\ \-\ \-\ \-\ \-\ \-\ \-\ u = \frac{2}{N}\sum_{i=1}^N \sin 2\phi_i
\label{eq:QU}
\end{equation}

The parameter $\mu$, known as the \textit{modulation factor}, represents the polarization fraction detected by the reconstruction algorithm for a 100\% polarized beam. A higher modulation factor indicates better performance of the reconstruction algorithm. The modulation factor is evaluated as:

\begin{equation}
\mu = \sqrt{\rm q^2 + u^2}.
\label{eq:modulation}
\end{equation}

While the modulation factor is specifically defined for a 100\% polarized source, the term \textit{modulation} will refer in this work to the same equation applied to sources with unknown polarization properties. We also define $\rm q_i = 2 \cos 2\phi_i$ and $\rm u_i = 2 \sin 2\phi_i$ as the Stokes parameters for each single detected photon, so that:

\begin{equation}
\rm {q = \frac{1}{N} \sum_{i=1}^N} q_i \-\ \-\ \-\ \-\ \-\ \-\ \-\ \-\ \-\ \rm{u = \frac{1}{N} \sum_{i=1}^N} u_i.
\end{equation}

In our previous work, we proposed a \textit{hybrid} analytic and machine learning (ML) approach to improve the performance of the standard algorithm \citep{cibrario}. By combining a Convolutional Neural Network (CNN) with the standard moment analysis, we achieved an improvement of the modulation factor and a significant mitigation of the \textit{polarization leakage} effect. It is extremely difficult to characterize and correct such systematics in the context of IXPE extended sources, due to the complex nature of their emission patterns. The hybrid method, by naturally mitigating this effect, offers a useful alternative to the standard moment analysis for analyzing these sources. In the previous work, which focused on the description of the algorithm and its feasibility study based on simulated data, polarization leakage was evaluated only in the context of simulated point sources \citep{cibrario}. 

Before applying a new reconstruction algorithm to scientific data, a validation phase of the method on experimental data is in order and crucial. The present work addresses this aspect describing the calibration phase of the algorithm conducted using real laboratory data. These are the official calibration data collected at the INAF-IAPS in Rome \citep{muleri}, and used for the calibration of IXPE detector units (DUs) before launch. The validation process of the algorithm involves measurements of both unpolarized and polarized laboratory beams, in order to characterize the instrumental systematics as well as to validate the reconstruction performance achieved with simulated data. This phase is critical, as previous CNN-based algorithms, that have been conceived for IXPE data reconstruction, and that achieved very promising results with simulated data \citep{Stanford_1, kitaguchi}, showed the biggest limitations in this context \citep{stanford_calibration}. The validation on laboratory data is needed not only to enhance robustness of the algorithm but also to bridge the gap between simulated environments and scientific data applications. In this work we report the first successful application of an ML-based algorithm for GPD track reconstruction to lab calibration data, marking a significant milestone in the feasibility study of ML techniques in this context. 

The successful validation of the algorithm with experimental data allowed us to apply the algorithm to IXPE observational data, and verifying its potentiality in the context of sources for which polarization leakage effect complicates the analysis and weakens the significance of the results. In this regard, we report the application of the algorithm to the analysis of an IXPE extended source, G21.5-0.9. This supernova remnant requires a complex analysis due to the significant impact of the polarization leakage, making it an ideal candidate to test the efficacy of our method in a practical setting. 

This paper is structured as follows. Section~\ref{sec:hybrid} provides an overview of the hybrid reconstruction method and of the results presented in \cite{cibrario}, reporting new findings related to simulated extended sources as well. Section~\ref{sec:calib} describes the calibration phase of the algorithm, depicting its application to real experimental data. Section \ref{sec:ixpe} briefly discusses the results of the G21.5-0.9 analysis. Finally, Section~\ref{sec:conclusions} summarizes our findings.

\section{An overview of the \textit{Hybrid method}}
\label{sec:hybrid}

A detailed description of the hybrid reconstruction algorithm can be found in \cite{cibrario}. Here, we only summarize the key characteristics and the main results, and we focus on the application of the method to astrophysical extended sources. The data set we used to develop the algorithm and to present the results discussed in the previous work and in this section consisted of Monte Carlo simulations, generated by using the \textit{ixpesim} software \citep{dilalla, ixpeobssim}. 

The algorithm exploits a CNN to predict the photon impact point in the GPD based on the photoelectron track image (see, e.g., Figure 1 of \cite{cibrario}). Our architecture is based on DenseNet-121 \citep{densenet} and incorporates hexagonal convolutional layers \citep{hexagdly}, implemented as a C++ extension for PyTorch.

\begin{figure}[htb]
    \centering
    \includegraphics[width=0.49\textwidth]{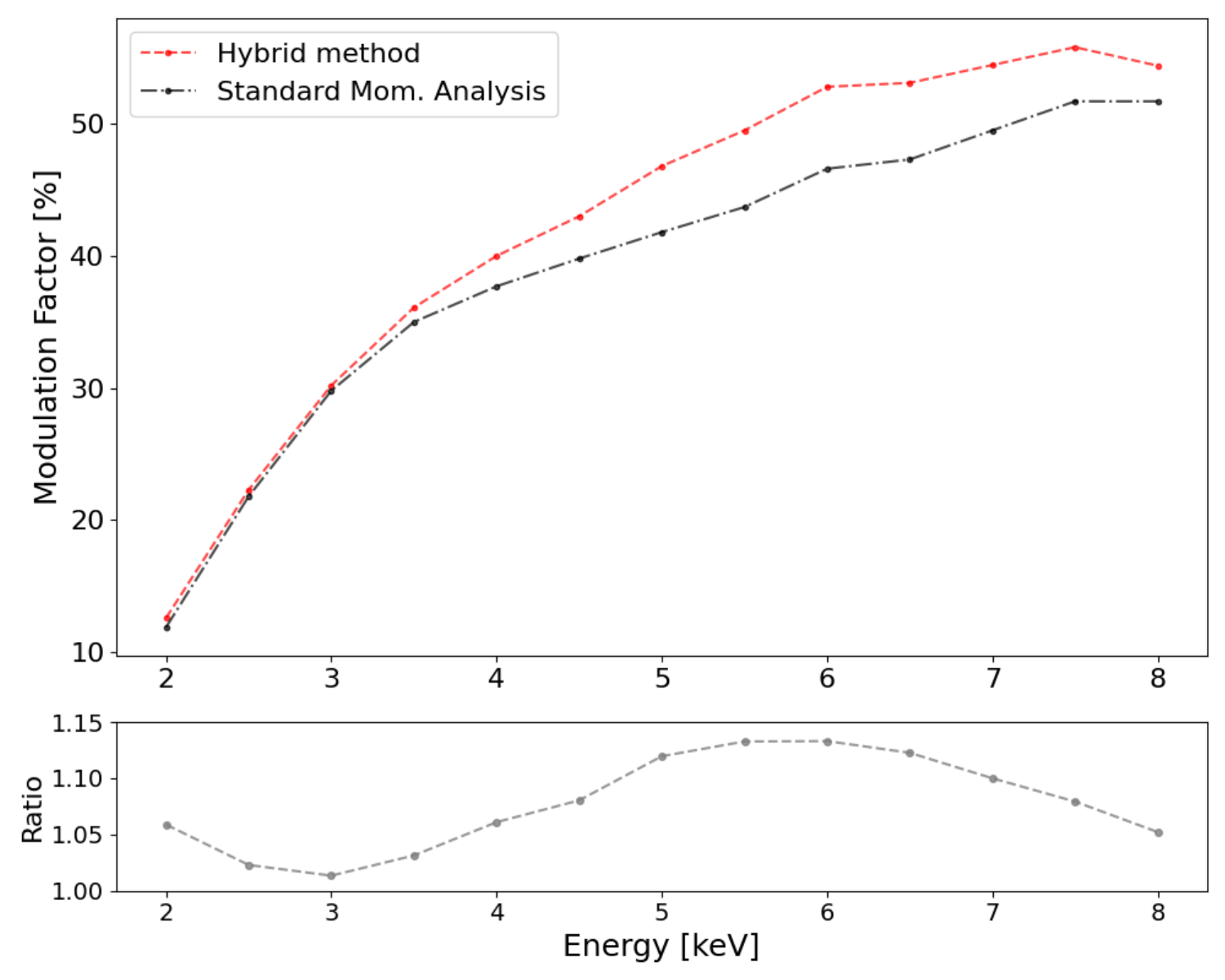}   
    \caption{\textit{Upper panel}: Modulation factor ($\mu$) as a function of energy achieved with simulated data, for the standard moment analysis in \textit{black} and for the hybrid method in \textit{red}. Error bars are present but not visible. \textit{Lower panel}: the ratio between the hybrid method $\mu$ and the standard one is reported as a \textit{gray dashed line}.}
    \label{fig:ip_mom_hyb}
\end{figure}

Our CNN demonstrated a significant improvement in the impact point reconstruction compared to the standard state-of-the-art analytic algorithm. Additionally, we applied an artificial sharpening process to the images, which further enhanced the performance. By integrating the CNN-predicted impact point into the moment analysis, we observed an improvement in the modulation factor, as illustrated in Fig.~\ref{fig:ip_mom_hyb}.

Our hybrid method also effectively reduced the systematic effect of the polarization leakage. In this regard it is important to note that polarization leakage mainly impacts the estimation of space-resolved polarization of extended sources, adding an artificial component perpendicular to the sharp edges of the source. The non-trivial geometry of emission areas complicates the handling and prediction of polarization leakage signal. There are significant efforts for developing a  correction of this effect by modeling a sky-calibrated Point Spread Function (PSF) of the instrument \citep{jack}. Polarization leakage has affected several IXPE analyses so far, including those of the Crab Nebula \citep{crab}, the Vela Pulsar Wind Nebula \citep{vela}, PSR B1509-58/MSH 15-52 \citep{hand_of_god}, and G21.5-0.9 (Di Lalla et al., in preparation), among others.

In our previous work we already demonstrated the ability of the algorithm to significantly mitigate the polarization leakage effect in the context of simulated point sources. Before validating the algorithm on experimental data, we studied the impact of the hybrid method in the mitigation of the polarization leakage with simulated extended sources as well. For this purpose, we generated a mock observation of 92 ks exposure—corresponding to the total on-source time of the February and March 2022 Crab Nebula IXPE observations—of an unpolarized source with intensity and properties similar to the Crab Nebula, as observed by a single IXPE detector unit. We make use of the \textit{ixpesim} simulation package (\textit{v14.3.5}) which allowed us to generate an IXPE observation based on a Chandra X-ray Observatory (CXO) dataset \citep{chandra}. For this test, we employed the CXO observation of the Crab Nebula (OBSID = 16364), without removing the readout streaks, to generate a photon list for use in the \textit{ixpesim} pipeline. The source was simulated without background contribution, employing the Point Spread Function (PSF) included in the v12 of the \textit{instrument response files (IRFs)}.
Being the simulated source unpolarized, every significant sign of residual polarization is due to the polarization leakage effect. This procedure is quite standard when analyzing extended sources, because it allows one to quantify the influence of the leakage on the observed polarization.

\begin{figure}[htb]
    \centering
    \includegraphics[width=0.499\textwidth]
    {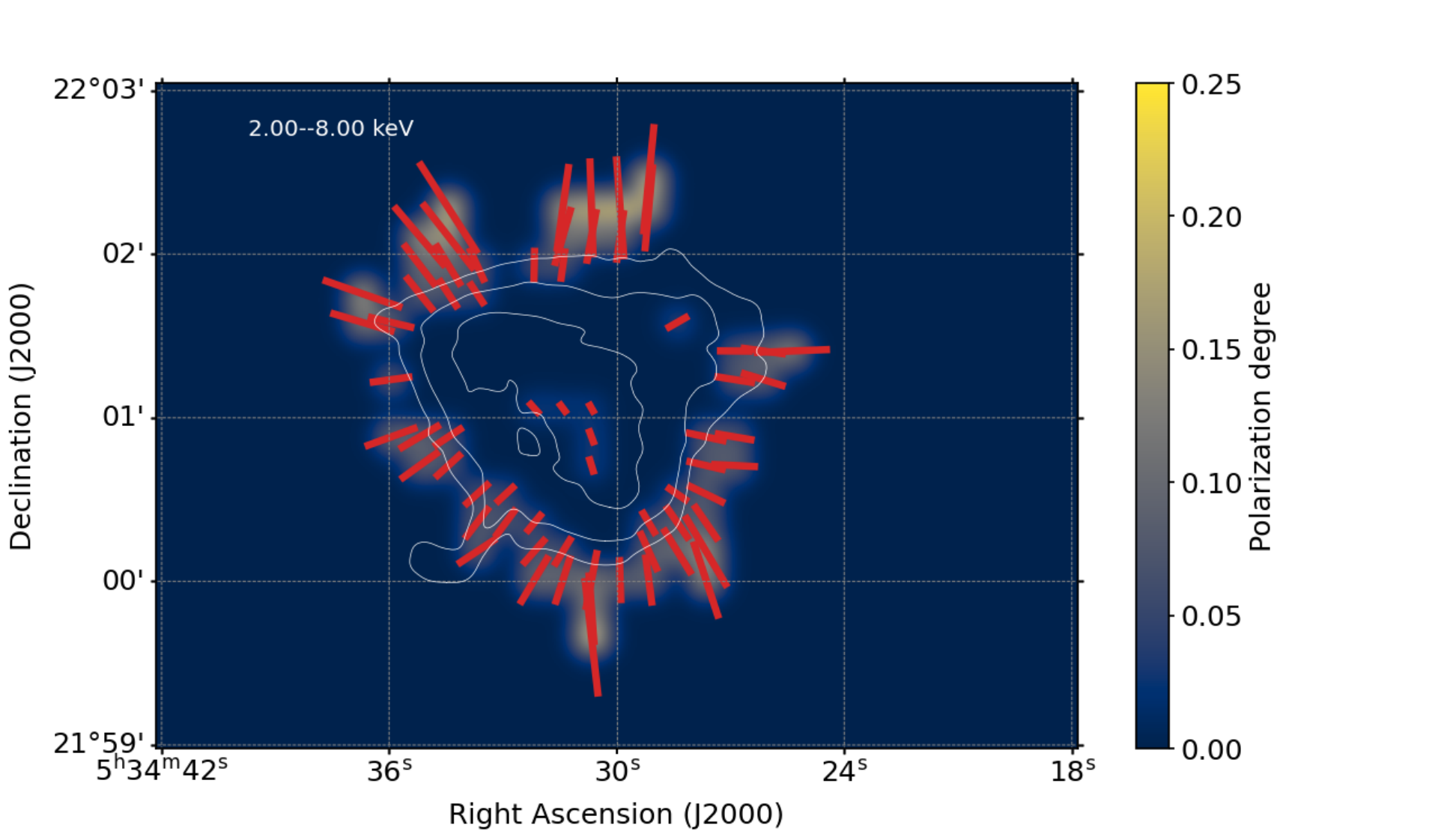}  
    \includegraphics[width=0.499\textwidth]
    {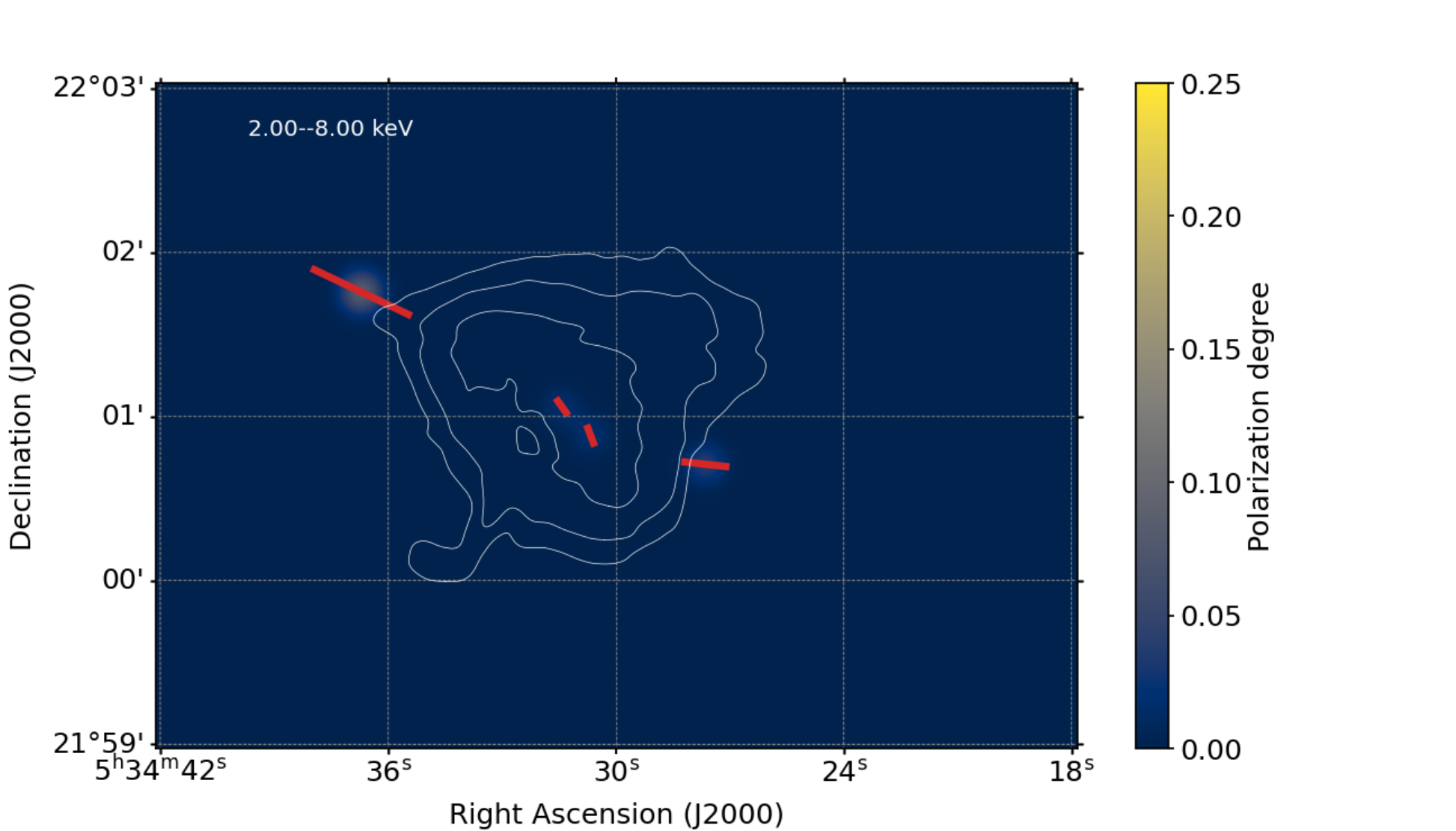}
    \caption{Polarization leakage effect on a simulated unpolarized extended source. \textit{Upper panel}: Polarization degree (color bar) and angle (red arrows) significantly ($>$3$\sigma$) detected by the standard moment analysis. \textit{Lower panel}: Polarization degree and angle significantly detected by the hybrid method. In both panels, the contours from the CXO observation are reported as \textit{white lines}.}
    \label{fig:crab_nebula}
\end{figure}

Fig.~\ref{fig:crab_nebula} reports the detected residual polarization degree and angle for the standard moment analysis in the upper panel, and for the hybrid method in the lower panel. Data were reduced using a standard tool of \textit{ixpeobssim} software (v13.0.1) \citep{ixpeobssim}. The maps are created by binning the data in the 2-8 keV energy range into a grid of $100 \times 100$ square pixels, with the application of a two-dimensional convolution of the Stokes maps using a $3 \times 3$ unit-filled kernel. Only the bins with a significant detection ($>$3$\sigma$) are represented. Examining the outcomes obtained with the standard moment analysis, the consequences of the polarization leakage effect are clearly detectable, with bins displaying a significant polarization degree up to approximately 25\% in the external region of the nebula. On the other hand, the mitigation of the polarization leakage achieved by the hybrid method has a substantial impact on the observation of the source. The lower panel features only four bins (instead of 67) where a significant measure of polarization is detected, with reduced intensities ($PD_{max} \sim$ 14\%). These results highlight the potential of this new algorithm on the observation of extended sources: the hybrid method could in fact enhance the significance of the signal, leading to a more robust detection.

\section{Validation of the algorithm with experimental data}
\label{sec:calib}

A rigorous and comprehensive validation of new methods conceived for the analysis of scientific data, such as IXPE data, is of utmost importance. This validation process consists in testing the algorithm with laboratory data, which provide the unique opportunity to control and arrange the actual energy and polarization properties of the measured radiation. For machine learning based algorithms this process is particularly important, as they could be very sensitive to the differences between simulations and real data \citep{stanford_calibration}. 

We employed the data acquired during the IXPE ground calibration campaign performed at INAF-IAPS in Rome, which were used to estimate the GPD performance and validate the standard moment analysis before IXPE launch. Throughout the calibration process, the three DUs were exposed to both unpolarized and polarized beams, spanning various energies in the 2-8 keV range. A detailed description of the laboratory setup and of the configurations used to generate the beams is available in \cite{muleri}. Here, in Table~\ref{tab:energies}, we list the energy values of the analyzed beams.

Unpolarized beams are used to evaluate the ability of the algorithm not to introduce biases in the measurement and can be employed to evaluate instrumental systematic effects. On the other hand, polarized radiation is used to measure the modulation factors at different input energies and to validate the resulting curve, shown for simulated data in Fig.~\ref{fig:ip_mom_hyb}.

\begin{table}[htb]
\caption{Summary of the energies of the X-ray unpolarized and polarized beams used to calibrate both the standard moment analysis and the hybrid method.}
\centering
\begin{tabular}{ll}
\hline
\hline
\noalign{\smallskip}
\textbf{Unpolarized beams} & \textbf{Polarized beams} \\
\noalign{\smallskip}
\textbf{energies [keV]} & \textbf{energies [keV]} \\
\hline
\noalign{\smallskip}
2.04 & 2.01 \\
\noalign{\smallskip}
2.29 & 2.29 \\
\noalign{\smallskip}
2.70 & 2.70 \\
\noalign{\smallskip}
2.98 & 2.98 \\
\noalign{\smallskip}
3.69 & 3.69\\
\noalign{\smallskip}
5.89 & 4.51 \\
\noalign{\smallskip}
 & 6.40 \\
\hline
\end{tabular}
\label{tab:energies}
\end{table}

\subsection{Unpolarized radiation}

When analyzing unpolarized beams, we still expect to measure a residual modulation due to two different effects. Firstly, only in a few cases it was possible to set-up calibration sources with a polarization lower than the sensitivity to achieve, so that a intrinsic source polarization was detectable in some cases. Additionally, an instrumental systematic effect known as \textit{spurious modulation} is also present \citep{gpd, Soffitta_2021}. The source of this systematic effect has not been conclusively identified yet, but it is thought to be due to an instrumental effect and it depends on the energy of the incident radiation and on the photon impact point position on the GPD plane. 

Characterizing this signal for all the IXPE DUs and across the whole energy interval (2-8 keV) is crucial in order to correct the actual observations for this spurious effect. This was already done for the analytical method before IXPE launch, and has to be now repeated with the hybrid method, as the responses of the two methods to this systematic effect could be different. Since the spurious modulation effectively introduces an additional contribution to the measured modulation, it must be separated from the true polarization of the beam source.

The work by \cite{sp_mod_rankin} establishes a method to disentangle the true polarization signal of the source from the spurious modulation. This method consists in conducting measurements by rotating the GPD around its center by 90$^\circ$, i.e. by orienting the GPD in 2 different directions perpendicular to each other. Under these conditions, from the GPD reference frame, the phase of the spurious modulation remains constant, while the phase of the intrinsic modulation generated by the source shifts by 90$^\circ$.

We analyzed the response of the three DUs to the unpolarized beams, one for each energy specified in Table~\ref{tab:energies}. The same energy beam was analyzed in the two configurations characterized by the 90$^\circ$ rotation (referred to as $\epsilon_1$ and $\epsilon_2$ in this work). In Table~\ref{tab:res_mod}, the measured modulation as a function of energy are reported for DU1. Similar results were achieved with DU2 and DU3 (see Appendix~\ref{app:a}). Note how the detected modulation for the different test sources varies with no obvious energy trend. This is expected due to both the energy dependence of the spurious modulation effect, and the intrinsic residual polarization which is unique to each energy beam source.

\begin{table}[htb]
\caption{Detected modulation (spurious modulation + intrinsic polarization) values obtained by analyzing unpolarized beams with DU1, in both $\epsilon_1$ and $\epsilon_2$ configuration, with the standard moment analysis and with the hybrid method.}
\centering
\footnotesize
\begin{tabular}{lllll}
\hline
\hline
\noalign{\smallskip}
\textbf{Energy [keV]} & \multicolumn{2}{l}{\textbf{Standard Mom.}} & \multicolumn{2}{l}{\textbf{Hybrid Method [\%]}} \\
\noalign{\smallskip}
 & \multicolumn{2}{l}{\textbf{Analysis [\%]}} & \\
 \hline
\noalign{\smallskip}
\textbf{DU1} & $\epsilon_1$ & $\epsilon_2$ & $\epsilon_1$ & $\epsilon_2$ \\
\hline
\noalign{\smallskip}
2.04 & $0.60 \pm 0.04$ & $0.69 \pm 0.04$ & $0.61 \pm 0.04$ & $0.71 \pm 0.04$ \\
\noalign{\smallskip}
2.29 & $0.54 \pm 0.04$ & $0.66 \pm 0.04$ & $0.54 \pm 0.04$ & $0.68 \pm 0.04$ \\
\noalign{\smallskip}
2.70 & $1.46 \pm 0.05$ & $1.78 \pm 0.05$ & $1.47 \pm 0.05$ & $1.79 \pm 0.05$ \\ 
\noalign{\smallskip}
2.98 & $4.21 \pm 0.05$ & $4.30 \pm 0.07$ & $4.32 \pm 0.05$ & $4.37 \pm 0.07$ \\
\noalign{\smallskip}
3.69 & $0.22 \pm 0.05$ & $0.30 \pm 0.05$ & $0.14 \pm 0.05$ & $0.22 \pm 0.05$ \\
\noalign{\smallskip}
5.89 & $0.12 \pm 0.05$ & $0.24 \pm 0.05$ & $0.15 \pm 0.05$ & $0.26 \pm 0.05$ \\
\hline
\end{tabular}
\label{tab:res_mod}
\end{table}

\begin{figure}[htb]
    \centering
    \includegraphics[width=0.33\textwidth]
    {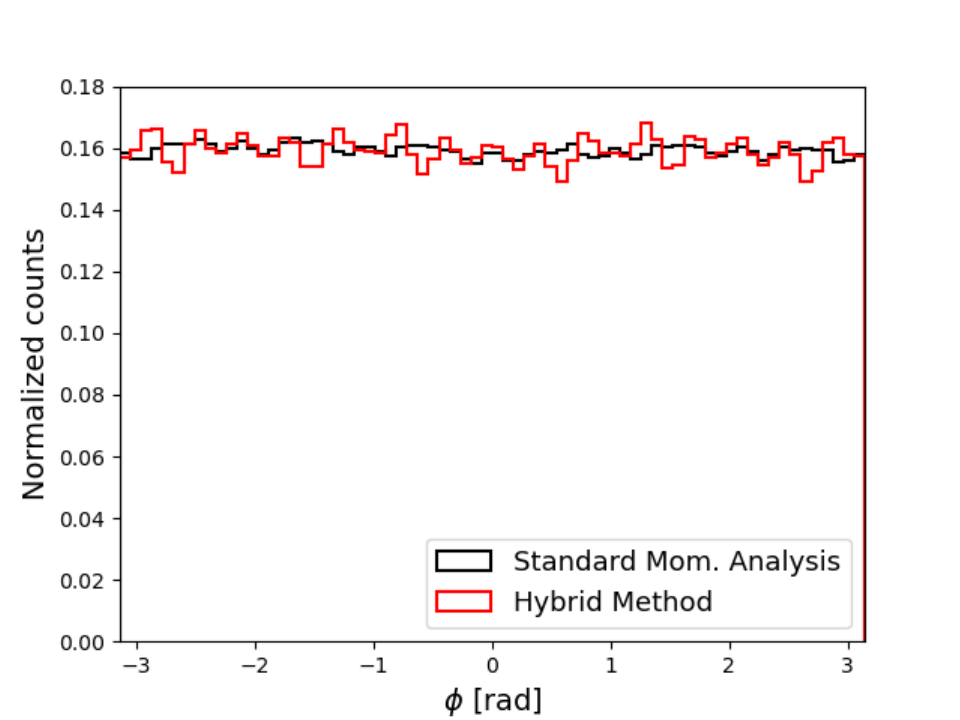}  
    \includegraphics[width=0.33\textwidth]
    {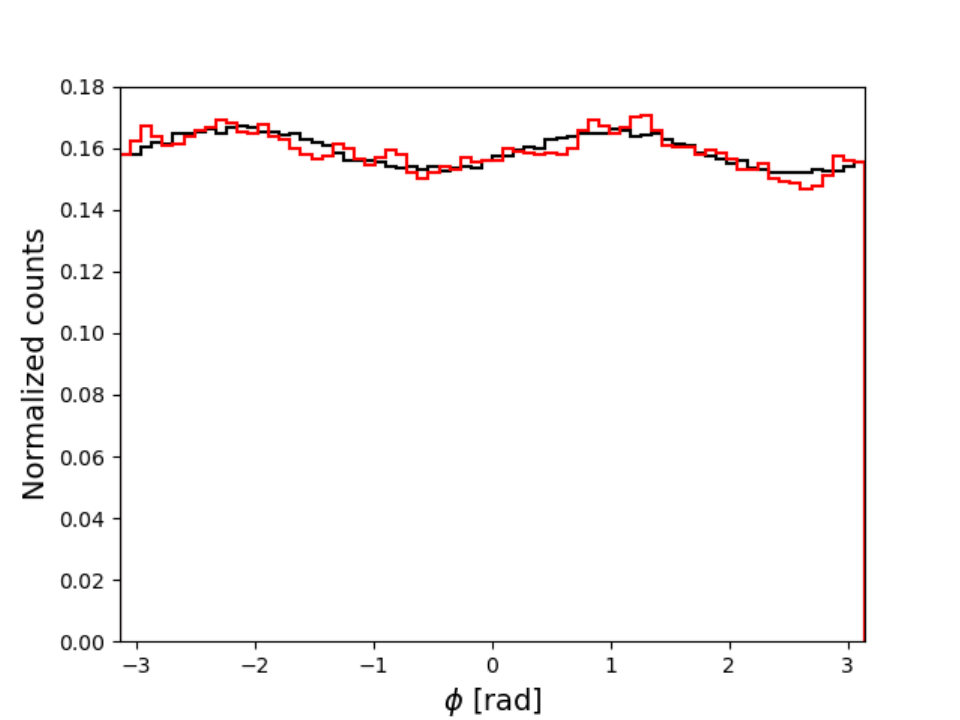}  
    \includegraphics[width=0.33\textwidth]
    {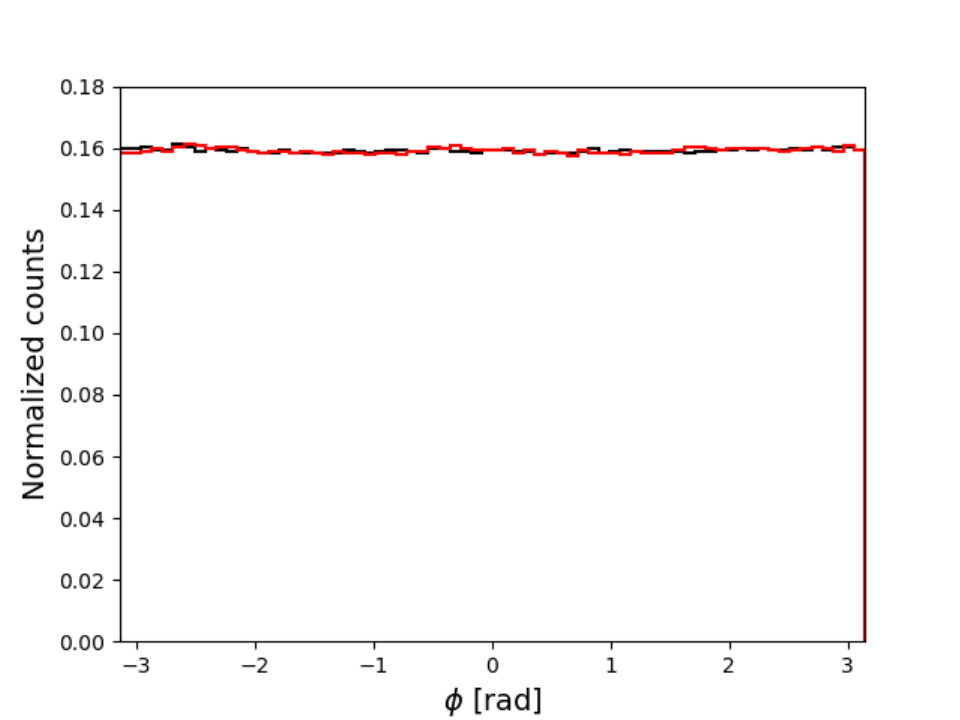}
    \caption{Emission angles distributions for three unpolarized energy beams, analyzed with DU1 at the $\epsilon_2$ configuration (\textit{Upper panel}: 2.04 keV; \textit{Middle panel}: 2.98 keV; \textit{Lower panel}: 5.89 keV). Results are reported for the hybrid method (\textit{red histogram}) and for the standard moment analysis (\textit{black histogram}).}
    \label{fig:mod_curves}
\end{figure}

In Fig.~\ref{fig:mod_curves}, as an example, the distributions of the predicted emission angles for three different unpolarized energy beams detected with DU1 and for the $\epsilon_2$ configuration are reported, for both the standard moment analysis in black and the hybrid method in red. The emission angles distribution shows a sinusoidal shape that is more pronounced at certain energies, while an ideal polarimeter would  measure only minimal modulation fluctuations compatible with zero, as seen in the simulated data for both reconstruction algorithms. As no correction to the data has been applied yet, the resulting modulation is due to the combined effect of the intrinsic beam polarization and the spurious modulation.

It should be noted that, apart from the mode $m = 2$, which arises from a combination of the beam polarization and the spurious modulation, higher-frequency minor peaks ($m = 6, 12$) appear in the emission angle distribution of the hybrid method for low energy beams (Fig.~\ref{fig:power_spec_27}). These peaks, which are less prominent in the distribution obtained using the standard moment analysis, were extensively investigated and understood not to be caused by the CNN (see Appendix~\ref{app:b}). Furthermore, they have no effect on the polarization results.

\begin{figure}[htb]
    \centering
    \includegraphics[width=0.49\textwidth]
    {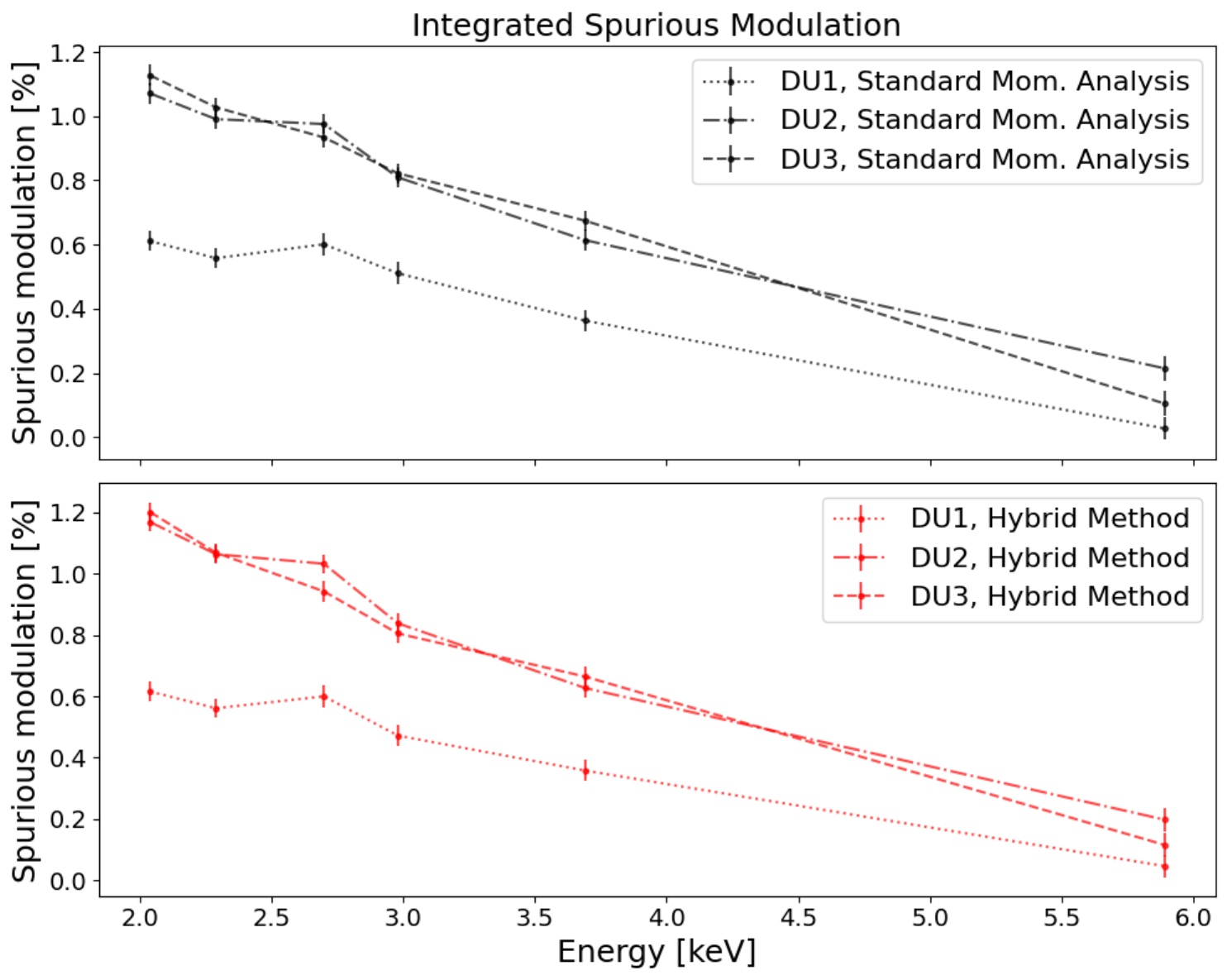} 
    \caption{Integrated spurious modulation for the central region (R $<$ 3 mm) of each IXPE DU for each energy beam. Results are reported for both the standard moment analysis (\textit{upper panel}) and the hybrid method (\textit{lower panel}).}
    \label{fig:sp_mod_int}
\end{figure}

We performed the same analysis described in \cite{sp_mod_rankin} to disentangle the true polarization signal of the source from the spurious modulation. Fig.~\ref{fig:sp_mod_int} shows the achieved spurious modulation signal integrated over the central region of the GPD (R $<$ 3 mm), for each DU and for both the standard moment analysis in black and the hybrid algorithm in red. 

\begin{figure*}[htb]
    \centering
    \includegraphics[width=0.4\textwidth]
    {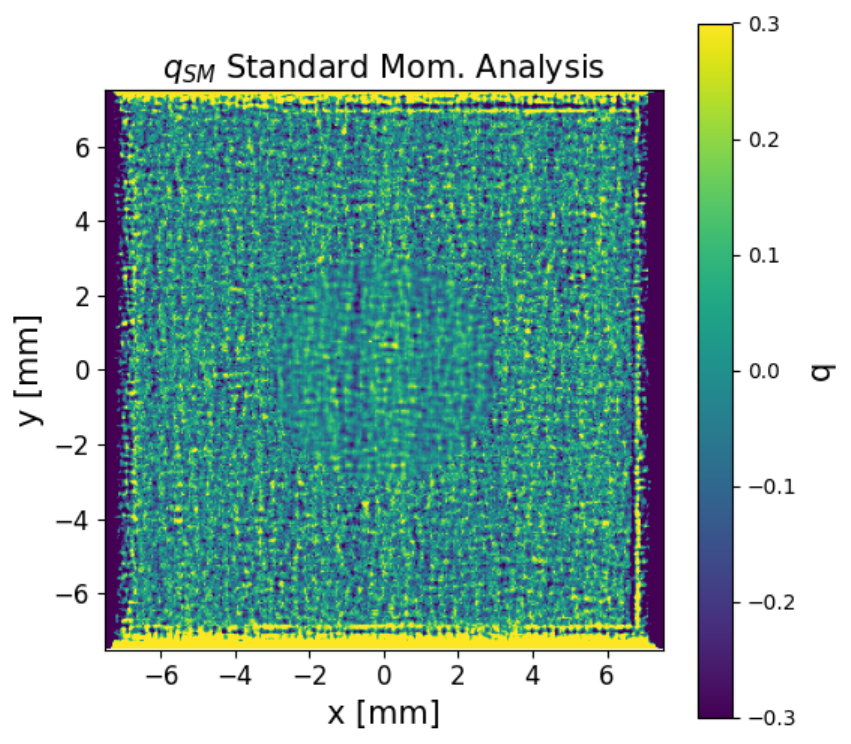}    
    \includegraphics[width=0.4\textwidth]
    {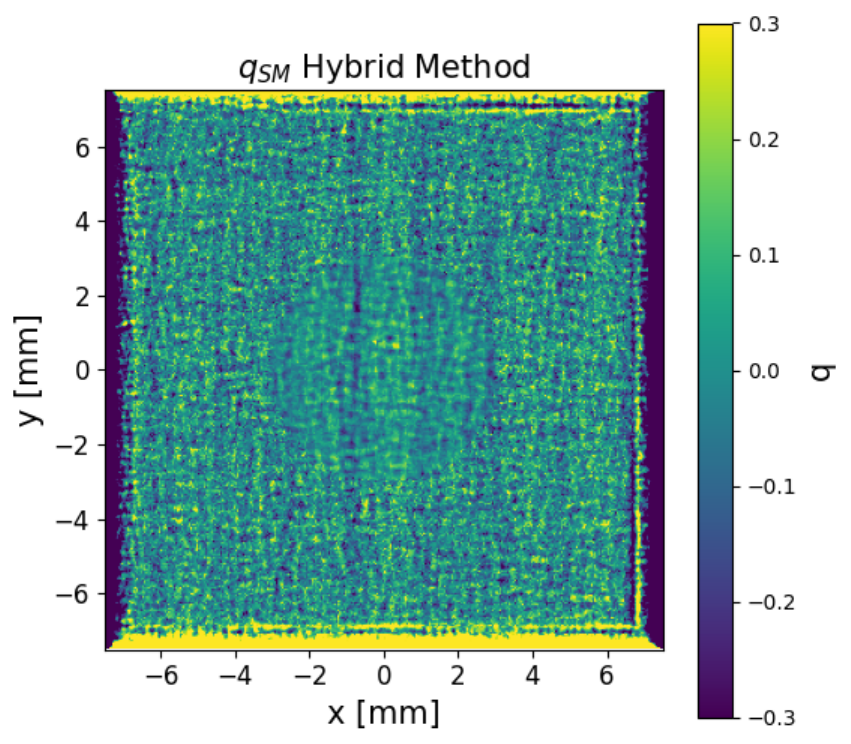}  
    \includegraphics[width=0.4\textwidth]
    {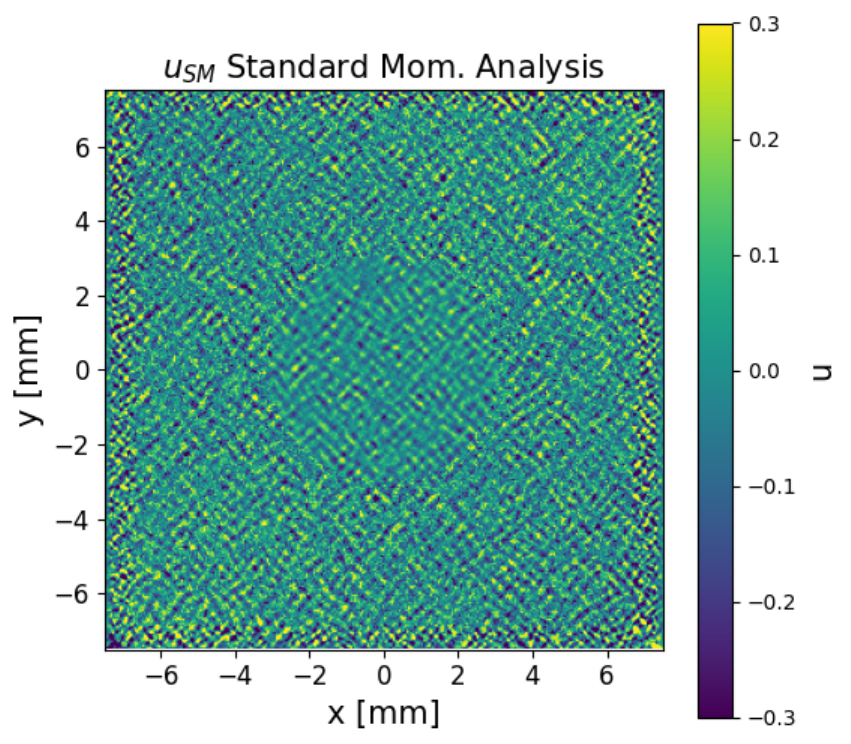}
    \includegraphics[width=0.4\textwidth]
    {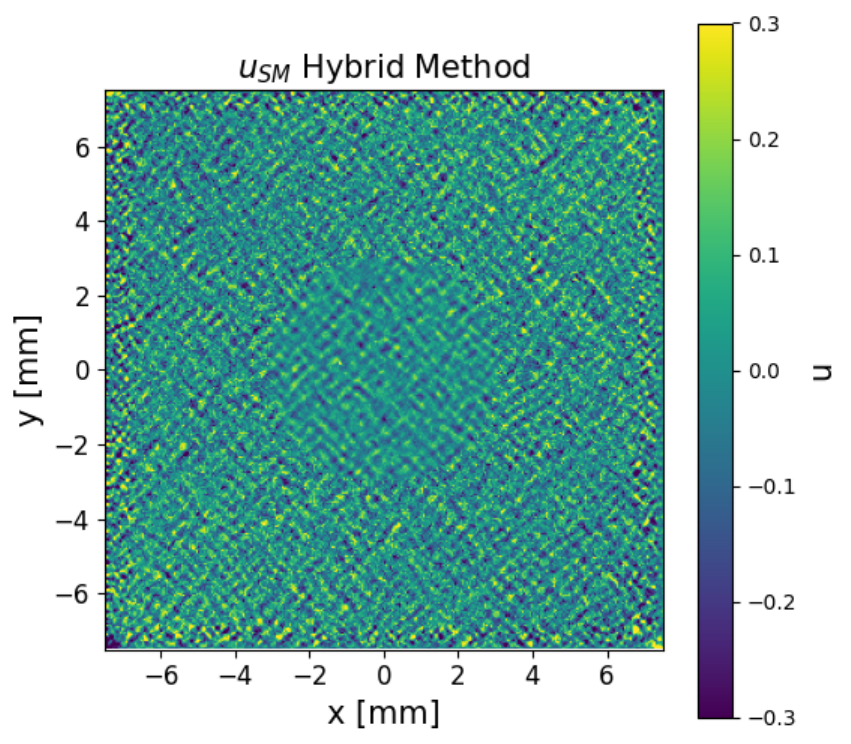} 
    \caption{Normalized Stokes parameters as a function of the impact point position on the GPD plane achieved by the standard moment analysis (\textit{left panels}) and by the hybrid method (\textit{right panels}) for a 2.7 keV unpolarized energy beam, for a single detector unit onboard IXPE (DU1).}
    \label{fig:QU_sp_maps}
\end{figure*}

As previously mentioned, the spurious modulation effect exhibits not only an energy dependence but also a position dependence. To account for this, following the procedure employed for the official IXPE calibration database (CALDB) spurious maps computation, we evaluated the spurious modulation maps for each energy beam, as illustrated in Fig.~\ref{fig:QU_sp_maps}. These maps report the normalized Stokes parameters (Eq.~\ref{eq:QU}), binned according to the photon impact point on the GPD plane, using a 300$\times$300 grid for binning. In the right panels of Fig.~\ref{fig:QU_sp_maps} the spurious modulation maps are reported for the 2.7 keV unpolarized beam, measured by IXPE DU1 and reconstructed with the hybrid method. As a reference, the same maps are reported in the left panels for the standard moment analysis. A clear residual polarization signal is detected in all the maps. The same maps are evaluated for all the sources at different energies and for all IXPE DUs.

By comparing the left and right panels of Fig.~\ref{fig:QU_sp_maps}, the similarity between the spurious modulation maps obtained through the standard moment analysis and the hybrid method is noteworthy, highlighting the presence of shared general structures. Our findings confirm that spurious modulation is a systematic caused by the detector, as it is observable with different reconstruction algorithms and differs for each DU.

Once these maps are produced, they are used to correct the q and u values obtained for each photon, accounting for the influence of spurious modulation on an event-by-event basis. The procedure corrects each photon Stokes parameters by subtracting the spurious modulation value relative to the position in the GPD where the photon is absorbed and to its measured energy, i.e. \citep{sp_mod_rankin}:

\begin{equation}
    \begin{aligned}
        \rm
        q_{corrected,i} &= \rm{q_{uncorrected,i} - q_{SM}[E][x][y]} \\
        \rm{u_{corrected,i}} &= \rm{u_{uncorrected,i} - u_{SM}[E][x][y]}
    \end{aligned}
\label{eq:sp_mod_correction}
\end{equation}
\noindent
where q(u)$_{\rm SM}$[E][x][y] are the normalized Stokes parameters of the spurious modulation maps corresponding to the photon energy E\footnote{Maps values are interpolated for events with energies which differ from the laboratory energy beams.} and impact point position (x,y).

\begin{figure}[htb]
    \centering
    \includegraphics[width=0.5\textwidth]
    {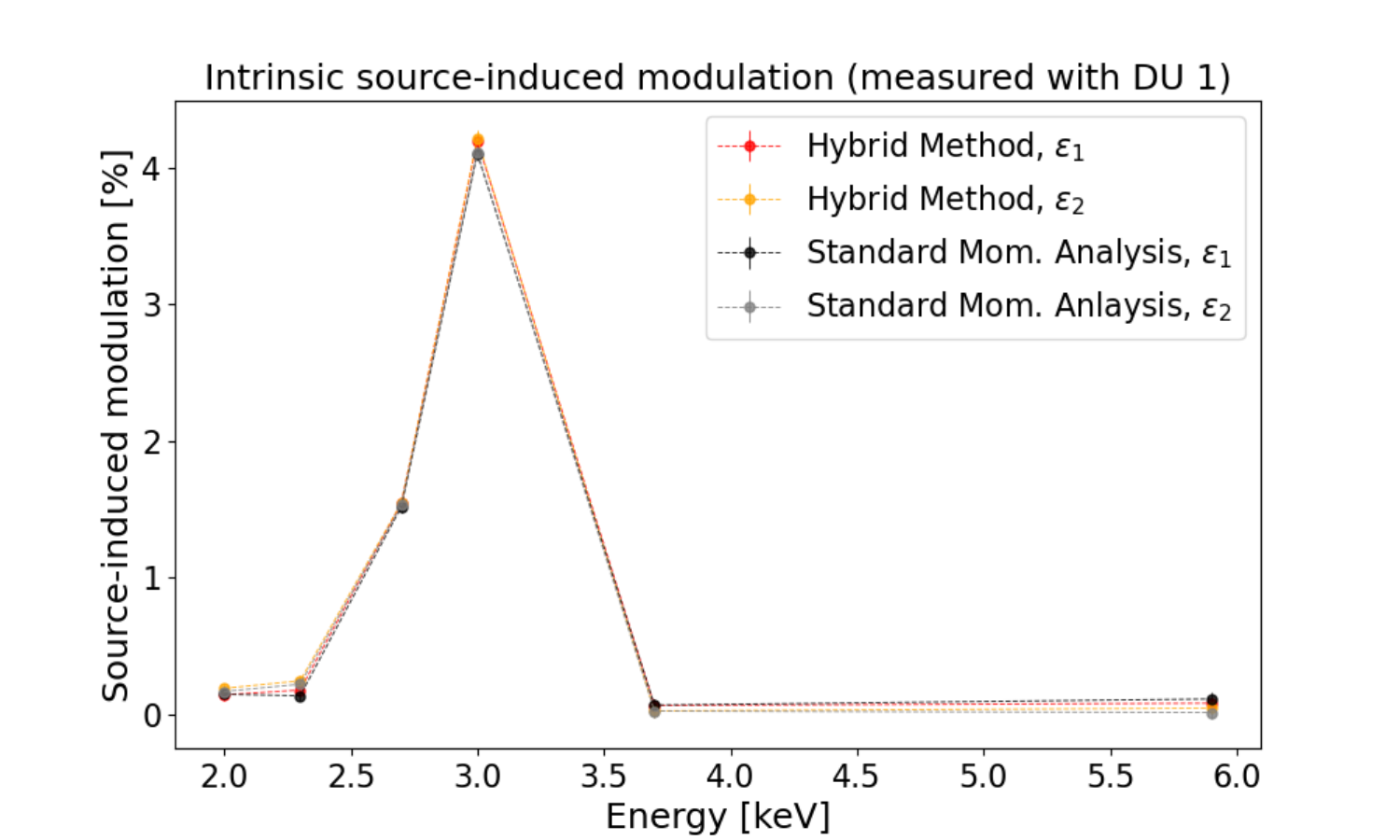}   
    \caption{Intrinsic source-induced modulation after the spurious modulation correction estimated with DU1 data. Results are reported for each unpolarized energy beam, for both the hybrid method ($\epsilon_1$ configuration in \textit{red}, $\epsilon_2$ configuration in \textit{orange}) and the standard moment analysis ($\epsilon_1$ configuration in \textit{black}, $\epsilon_2$ configuration in \textit{gray}).}
    \label{fig:residual_m_du1}
\end{figure}

It is important to validate the procedure by confirming that the intrinsic source-induced modulation, estimated by individually correcting the unpolarized beams results, is compatible for the $\epsilon_1$ and $\epsilon_2$ configurations. In Fig.~\ref{fig:residual_m_du1} the comparison between the intrinsic source-induced modulation estimated for both the configurations and methods are reported for DU1: all the results are compatible within the error bars, validating the procedure of the spurious modulation subtraction. The same compatibility was achieved for DU2 and DU3, and compatible results were achieved across DUs as well (see Appendix~\ref{app:a}).

All the results presented in this section confirm the proper functioning of the hybrid method when used for analyzing unpolarized radiation in laboratory. The spurious maps closely resemble the ones obtained with the standard moment analysis, the sources-induced modulation is comparable, and no significant bias or unexpected behavior is detected.

\subsection{Polarized radiation}

\begin{figure*}
\includegraphics[width=0.33\textwidth]{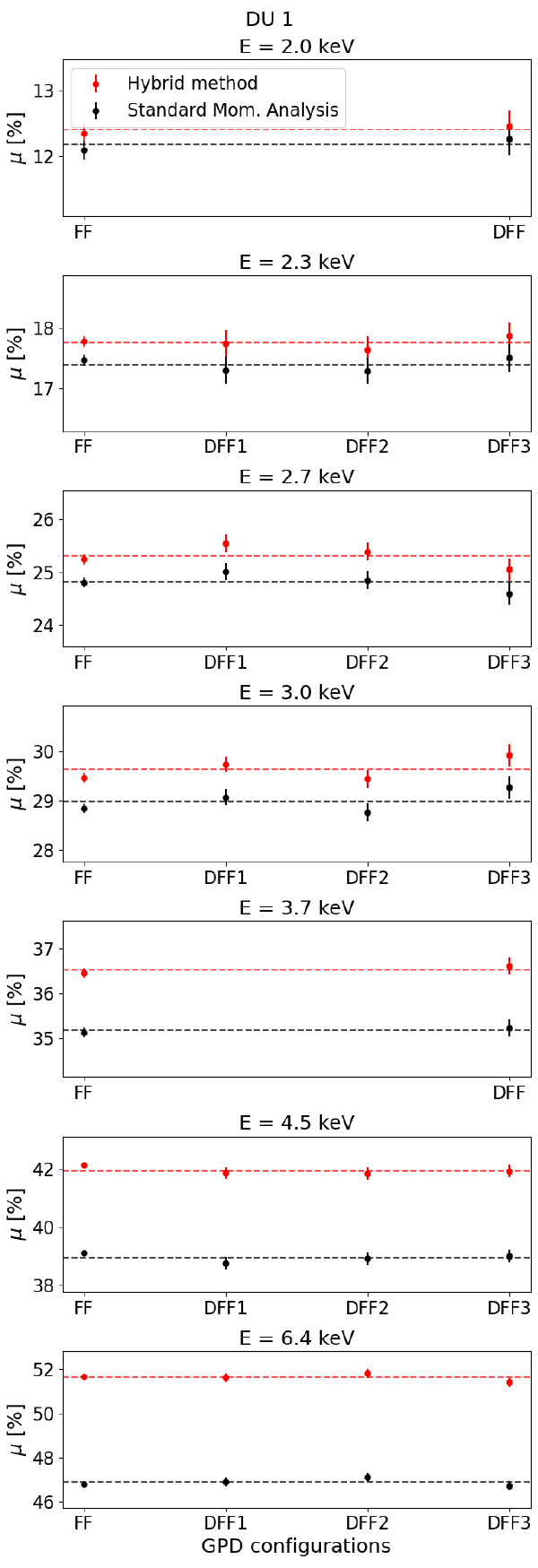}
\includegraphics[width=0.33\textwidth]{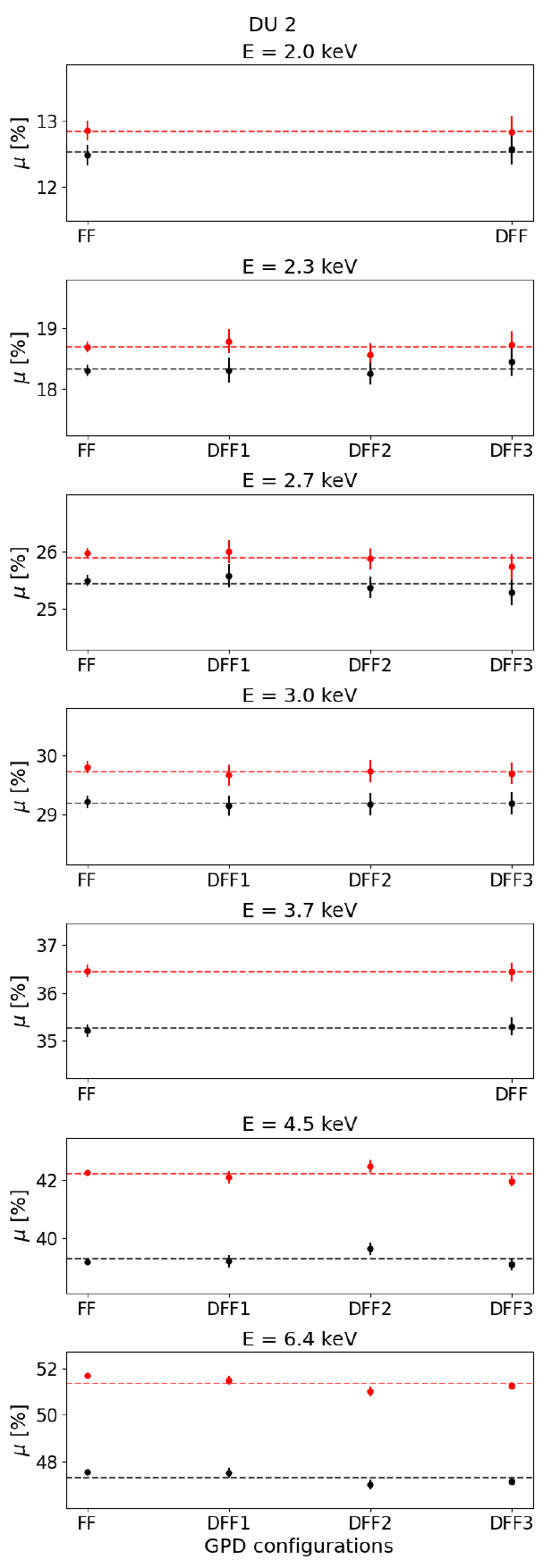}
\includegraphics[width=0.33\textwidth]{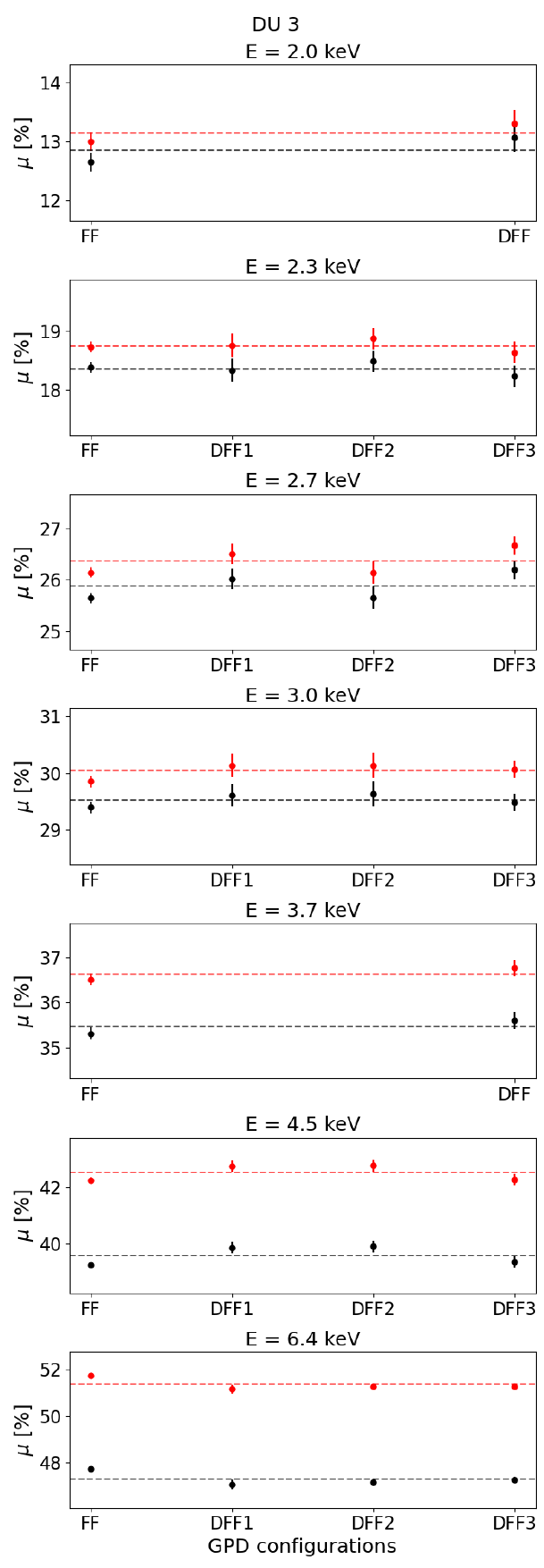}
\caption{Modulation factor ($\mu$) achieved by the hybrid method (\textit{red}) and standard moment analysis (\textit{black}) for all DUs onboard of IXPE, and for each polarized energy beam and GPD configuration (FF or DFF).}
\label{fig:mod_allE}
\end{figure*}

Once the unpolarized beams analysis was completed, we moved to the polarized data set analysis. The methodology applied to analyze polarized beams is equivalent to the one used with unpolarized beams: each event is corrected by the spurious modulation effect following Eq.~\ref{eq:sp_mod_correction}.

The GPD's response to polarized radiation could, in principle, exhibit spatial dependence (similar to the spurious modulation effect) and phase dependence. During the IXPE calibration campaign, the spatial and phase uniformity of the GPD's response to polarized radiation, as determined using the standard moment analysis, was thoroughly analyzed and confirmed \citep{mod_factor_calib}. Any potential non-uniformity was hypothesized to arise from instrumental effects, rather than from the reconstruction algorithm itself. Nevertheless, for completeness, we analyzed the response of IXPE GPDs to polarized radiation for each energy beam under two or, when available, four different configurations: a Flat Field (FF) configuration, where the entire surface of the GPD is exposed to incident radiation, and three distinct Deep Flat Field (DFF) configurations, each defined by a different polarization angle, where only the central region of the GPD (R $<$ 3.3 mm) is exposed to incident radiation.

The estimated modulation factor for each IXPE DU and for each energy and phase configuration is reported in Fig.~\ref{fig:mod_allE}, for both the standard moment analysis and the hybrid method. The final modulation factor for each energy beam is evaluated as the mean value of the $\mu$s detected in the two (or four) configurations, and its uncertainty is calculated as the semi-dispersion of the measures. 

\begin{figure}
\centering
\includegraphics[width=0.45\textwidth]{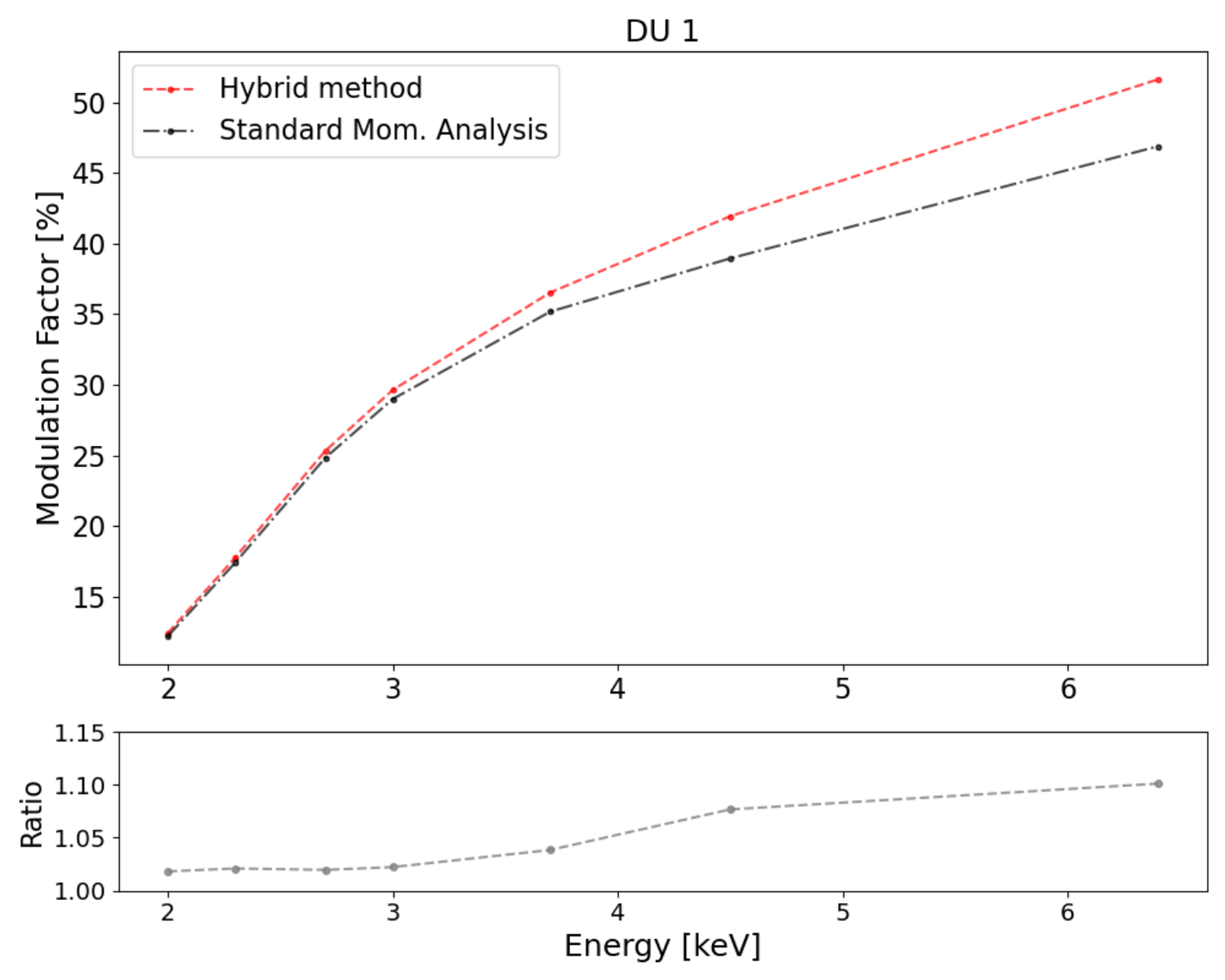}
\includegraphics[width=0.45\textwidth]{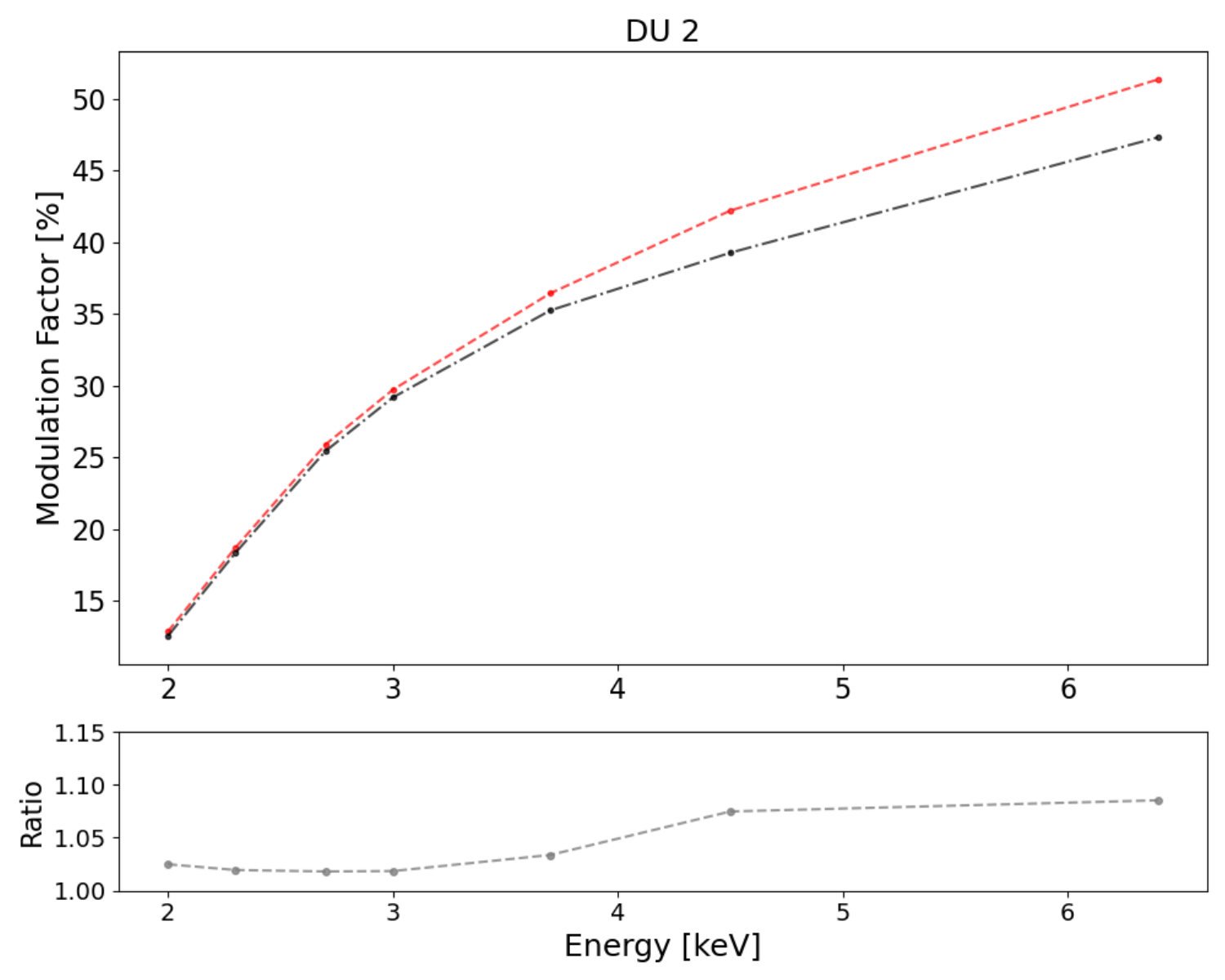}
\includegraphics[width=0.45\textwidth]{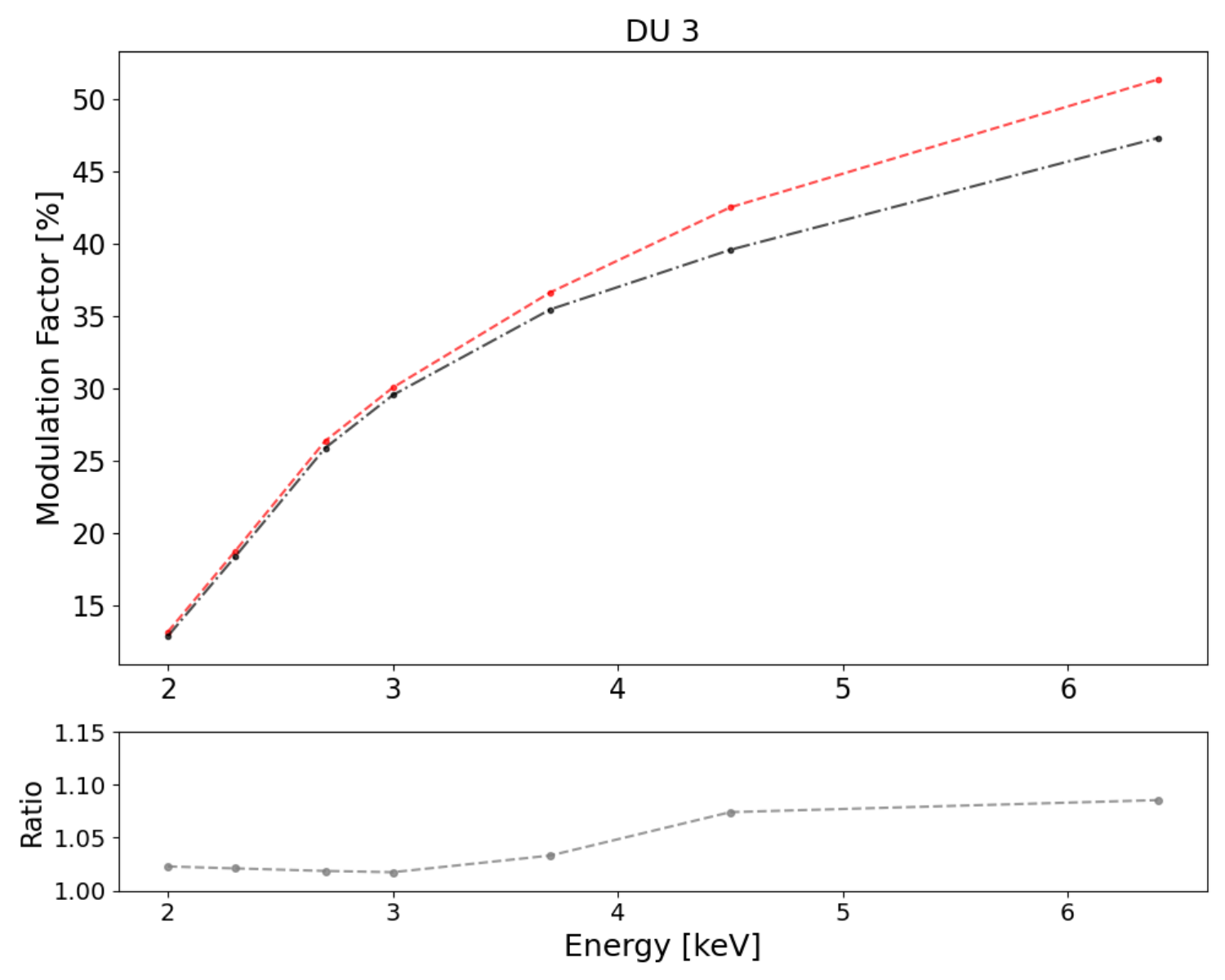}
\caption{Modulation factor curves as a function of the energy achieved by the hybrid method (\textit{red}) and standard moment analysis (\textit{black}) for all DUs onboard of IXPE. Error bars are present but not visible. The ratios between the hybrid method $\mu$ and the standard one are reported for each DU.}
\label{fig:final_mod}
\end{figure}

The final modulation factors as a function of the energy for the three DUs are reported in the three panels of Fig.~\ref{fig:final_mod}. The comparison between the standard moment analysis and the hybrid method closely resembles the one obtained with simulations in Fig.~\ref{fig:ip_mom_hyb}. The improvement is indeed marginal up to 3 keV, it is $\sim$1.5\% at 3.7 keV, and peaks at $\sim$5\% at 6.4 keV. 

The pressure within the GPD gas cell is not constant over time and decreases gradually \citep{IXPE_prelaunch}. This pressure variation impacts the modulation factor, as it slightly alters the length and shape of the tracks, contributing to the small differences observed in the $\mu$ values across different DUs, as well as between calibration data and simulations. By indirectly monitoring the temporal pressure changes in the GPD, we can adjust the modulation factor curves to align with the observation period of the specific IXPE source under analysis. For the hybrid method, three distinct modulation factor curves are generated for each DU, corresponding to each year of IXPE's observational campaign. This approach follows the one used by the collaboration for the standard moment analysis.

\section{Application to IXPE data}
\label{sec:ixpe}

Following the calibration phase of the algorithm, all the necessary components to conduct IXPE sources data analysis using the hybrid method are now available. As outlined in previous sections, the polarization leakage effect is a significant systematic uncertainty in observations of extended sources, particularly in regions with sharp intensity gradients and edges. As already mentioned in Sec.~\ref{sec:hybrid}, we observe a substantial reduction of the polarization leakage effect when applying the hybrid method to reconstruct the photoelectron tracks of simulated data (see Fig.~\ref{fig:crab_nebula}). 

\begin{figure}[htb]
    \centering
    \includegraphics[width=0.499\textwidth]{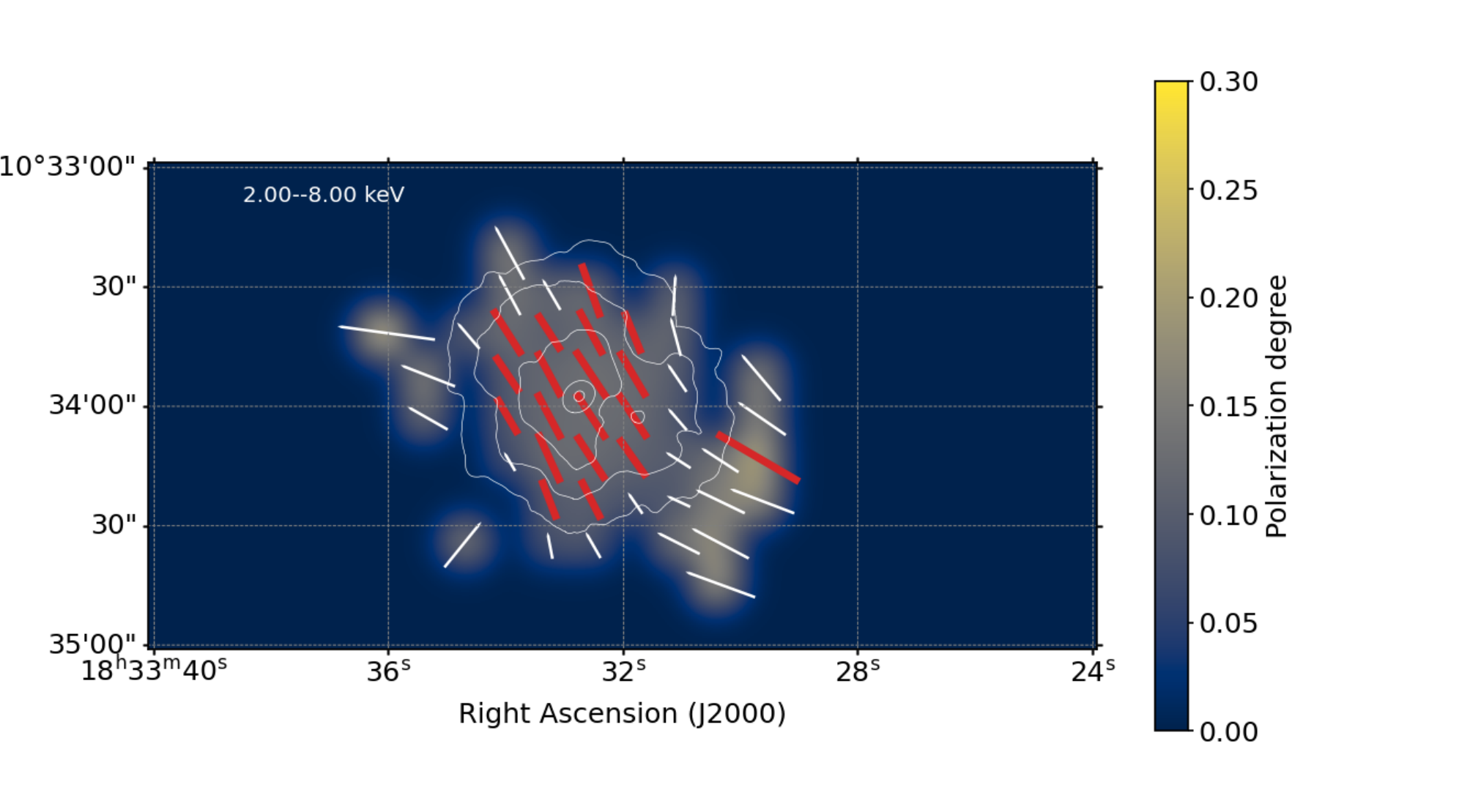} 
    \includegraphics[width=0.499\textwidth]{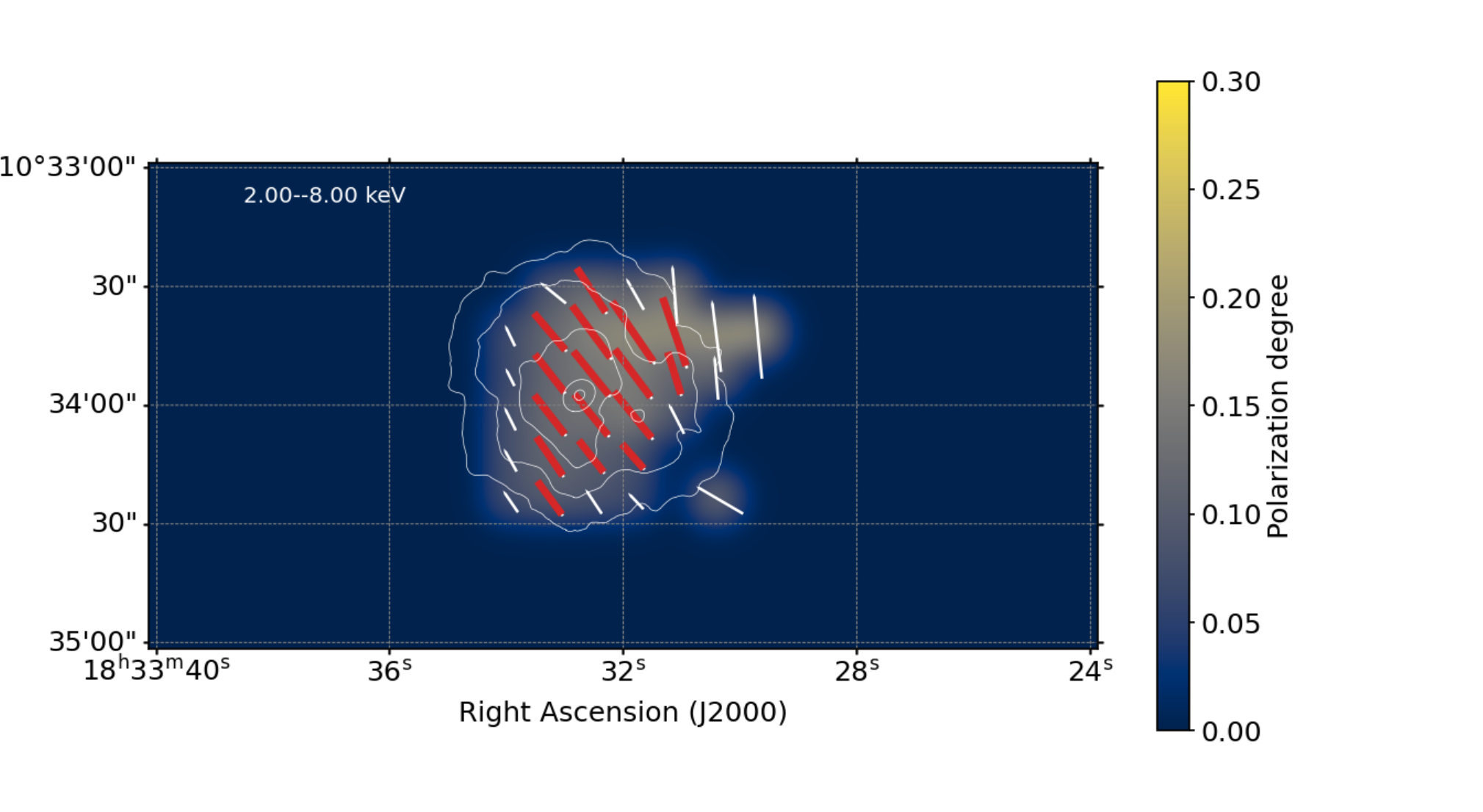}
    \caption{Binned polarization maps from the IXPE observation of G21.5-0.9, generated using the standard moment analysis (\textit{upper panel}) and the hybrid method (\textit{lower panel}). The arrows orientation represents the polarization angle, and the color map corresponds to the polarization degree. Bins with 2$\sigma$ significance are marked in \textit{white}, while those with 3$\sigma$ significance are highlighted in \textit{red}. In both panels, the contours from the CXO observation (OBSID: 01433) are reported as \textit{white lines}.}
    \label{fig:g21}
\end{figure}

In this section, we present an example of the improvement on experimental IXPE data, specifically for G21.5-0.9, a supernova remnant observed by IXPE in October 2023, with a total livetime of approximately 840 ks (OBSID: 02001199). We processed the IXPE data from this observation using both the standard moment analysis and the hybrid method, comparing the resulting polarization maps, as shown in Fig.~\ref{fig:g21}.

The same binning and convolution procedure described in Sec.~\ref{sec:hybrid} for the unpolarized Crab Nebula was applied for this source as well. A marked difference in the structure of the polarization maps is evident when comparing the two methods, particularly in the outer regions of the nebula, where the influence of the polarization leakage effect is expected to be prominent. These outer regions are crucial for studying the magnetic field structures responsible for the X-ray emission and assessing their potential alignment with radio observations of the source. The differences observed using the two different methods are critical in this regard, and have significant implications on the understanding of the physical processes governing the remnant's emission. While a discussion of the astrophysical implications lies beyond the scope of this paper, a detailed analysis will be presented in separate study (Di Lalla et al., in preparation).

\section{Conclusions and future prospects}
\label{sec:conclusions}

X-ray polarimetric imaging provides crucial insights into the magnetic fields geometry and emission mechanisms of astrophysical sources, by enabling spatially resolved polarization measurements that reveal patterns across extended regions. Polarization leakage, a systematic effect caused by inaccuracies in reconstructing the photon impact point in the photoelectron tracks, can induce spurious polarization patterns linked to intensity gradients, affecting our understanding of the true polarization properties of these sources. Developing and applying mitigation techniques is essential to address polarization leakage, ensuring the reliability of polarimetric measurements and the scientific interpretations derived from them.

In this work, we presented the first successful application of a ML-based algorithm (described in \cite{cibrario}) to experimental lab data, which further confirmed the algorithm potential. Measurements of unpolarized beams revealed no unexpected residual polarization after correcting for spurious modulation, a critical outcome for ML-based algorithms given their sensitivity to such biases. This key result underscores the robustness of the approach, confirming the absence of systematic biases. Moreover, the modulation factors for 100\% polarized beams align with simulation trends, consolidating the algorithm reliability with experimental data. The application of the hybrid method to an IXPE extended source, G21.5-0.9, showed a substantial improvement over the current standard analysis method in terms of polarization leakage, with significant implications on the physical interpretation of the data.

In summary, our hybrid analytic-ML approach offers advancements in GPD-based X-ray polarimetry, enhancing performance, mitigating the systematic effect of polarization leakage, and demonstrating robustness with experimental data. While initially conceived for future X-ray polarimetry missions, these results demonstrated the value of this new approach also for the current measurements with  IXPE mission, providing a practical framework for future advancements in X-ray polarimetry. Notably, all IXPE observations can be potentially reprocessed with the hybrid method.

We encourage anyone interested in using the code -- whether for data analysis with the hybrid reconstruction pipeline or for exploring specific components, such as the network architecture, for applications outside the field -- to contact the authors. We are happy to provide guidance and collaborate to facilitate its use for a wide range of purposes.

\begin{acknowledgements} 
Portions of this research were conducted with high performance computing resources provided by Louisiana State University (http://www.hpc.lsu.edu).
We acknowledge Federica Legger, Sara Vallero, and the INFN Computing Center of Turin for providing support and computational resources. This work is part of the project “SKYNET: Deep Learning for Astroparticle Physics”, PRIN 2022 (CUP: D53D23002610006), funded by the European Union – Next Generation EU, Mission 4 Component 1. This work is part of the NODES project, funded by the MUR – M4C2 1.5 of PNRR with grant agreement no. ECS00000036.
\end{acknowledgements}

\bibliography{bibliography}{}
\bibliographystyle{aasjournal}

\begin{appendix}

\section{Additional results from DU2 and DU3}
\label{app:a}

In the main text, we primarily presented results obtained from DU1 data. Table~\ref{tab:res_mod} showed the modulation measured for each energy beam and configuration before any corrections for systematic effects, thus including the spurious modulation component. For comparison, corresponding values for DU2 and DU3 are presented in Table~\ref{tab:res_mod_DU23}. Given that these values are affected by instrumental effects, variations across DUs are expected in these results.

\begin{table}[htb]
\caption{Residual modulation values obtained by analyzing unpolarized beams with DU2 and DU3 (in both $\epsilon_1$ and $\epsilon_2$ configuration) with the standard moment analysis and with the hybrid method.}
\centering
\footnotesize
\begin{minipage}{0.48\textwidth}
\centering
\begin{tabular}{lllll}
\hline
\hline
\noalign{\smallskip}
\textbf{Energy [keV]} & \multicolumn{2}{l}{\textbf{Standard Mom.}} & \multicolumn{2}{l}{\textbf{Hybrid Method [\%]}} \\
\noalign{\smallskip}
 & \multicolumn{2}{l}{\textbf{Analysis [\%]}} & \\
 \hline
\noalign{\smallskip}
\textbf{DU2} & $\epsilon_1$ & $\epsilon_2$ & $\epsilon_1$ & $\epsilon_2$ \\
\hline
\noalign{\smallskip}
2.04 & $1.03 \pm 0.04$ & $0.88 \pm 0.04$ & $1.12 \pm 0.04$ & $0.54 \pm 0.04$ \\
\noalign{\smallskip}
2.29 & $0.93 \pm 0.04$ & $0.72 \pm 0.04$ & $0.99 \pm 0.04$ & $0.79 \pm 0.04$ \\
\noalign{\smallskip}
2.70 & $2.28 \pm 0.05$ & $1.16 \pm 0.05$ & $2.37 \pm 0.05$ & $1.17 \pm 0.05$ \\ 
\noalign{\smallskip}
2.98 & $4.65 \pm 0.05$ & $3.61 \pm 0.05$ & $4.75 \pm 0.05$ & $3.69 \pm 0.05$ \\
\noalign{\smallskip}
3.69 & $0.59 \pm 0.05$ & $0.54 \pm 0.05$ & $0.55 \pm 0.05$ & $0.51 \pm 0.05$ \\
\noalign{\smallskip}
5.89 & $0.18 \pm 0.05$ & $0.05 \pm 0.05$ & $0.17 \pm 0.05$ & $0.05 \pm 0.05$ \\
\hline
\end{tabular}
\end{minipage}%
\hfill
\begin{minipage}{0.48\textwidth}
\centering
\begin{tabular}{lllll}
\hline
\hline
\noalign{\smallskip}
\textbf{Energy [keV]} & \multicolumn{2}{l}{\textbf{Standard Mom.}} & \multicolumn{2}{l}{\textbf{Hybrid Method [\%]}} \\
\noalign{\smallskip}
 & \multicolumn{2}{l}{\textbf{Analysis [\%]}} & \\
 \hline
\noalign{\smallskip}
\textbf{DU3} & $\epsilon_1$ & $\epsilon_2$ & $\epsilon_1$ & $\epsilon_2$ \\
\hline
\noalign{\smallskip}
2.04 & $1.04 \pm 0.04$ & $0.66 \pm 0.04$ & $0.91 \pm 0.04$ & $0.71 \pm 0.04$ \\
\noalign{\smallskip}
2.29 & $0.88 \pm 0.04$ & $0.69 \pm 0.04$ & $0.91 \pm 0.04$ & $0.71 \pm 0.04$ \\
\noalign{\smallskip}
2.70 & $2.22 \pm 0.05$ & $1.27 \pm 0.05$ & $2.27 \pm 0.05$ & $1.32 \pm 0.05$ \\ 
\noalign{\smallskip}
2.98 & $4.82 \pm 0.05$ & $4.09 \pm 0.05$ & $4.89 \pm 0.05$ & $4.18 \pm 0.05$ \\
\noalign{\smallskip}
3.69 & $0.42 \pm 0.05$ & $0.33 \pm 0.05$ & $0.38 \pm 0.05$ & $0.27 \pm 0.05$ \\
\noalign{\smallskip}
5.89 & $0.17 \pm 0.05$ & $0.17 \pm 0.05$ & $0.18 \pm 0.05$ & $0.16 \pm 0.05$ \\
\hline
\end{tabular}
\end{minipage}
\label{tab:res_mod_DU23}
\end{table}

\begin{figure}[htb]
    \centering
    \includegraphics[width=0.49\textwidth]{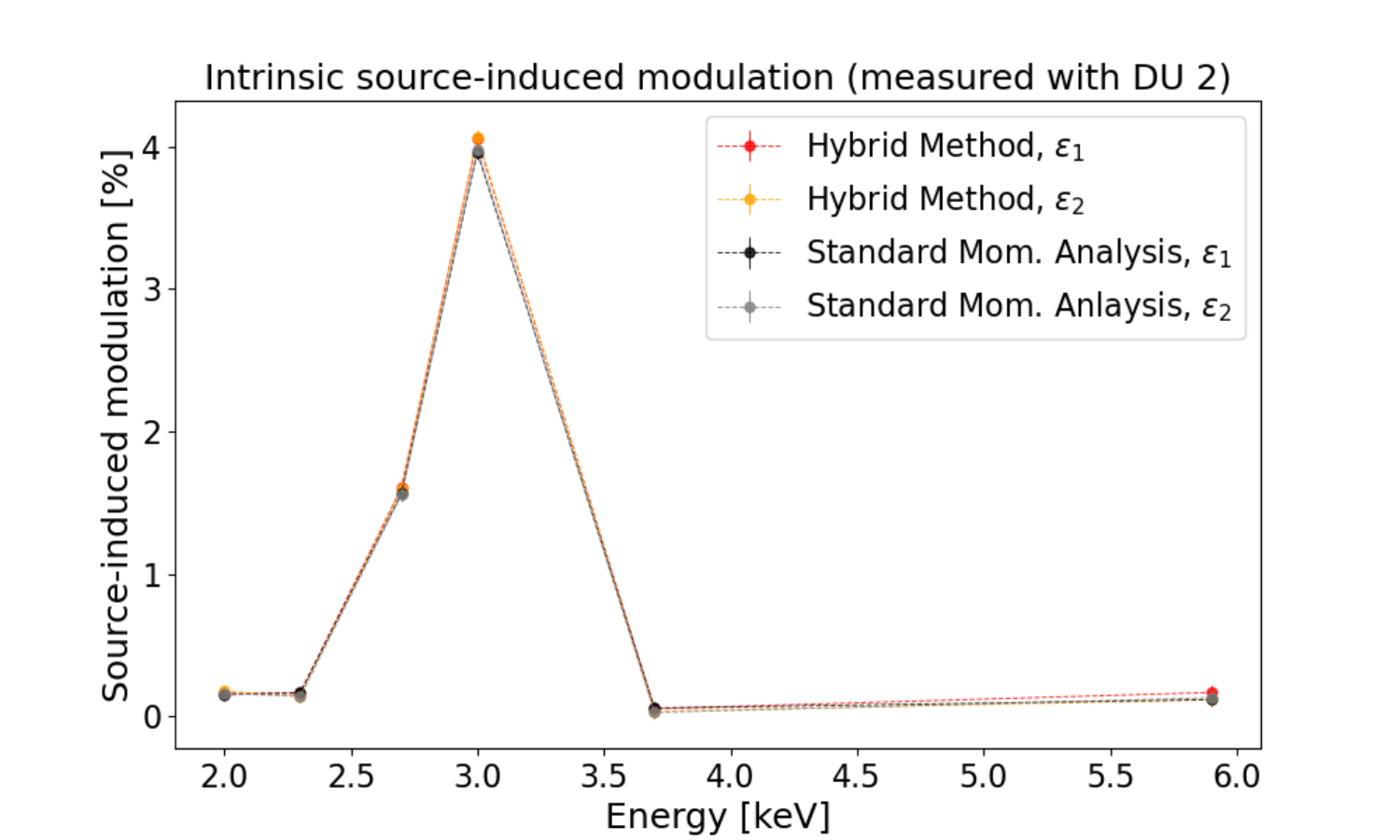}   
    \includegraphics[width=0.49\textwidth]{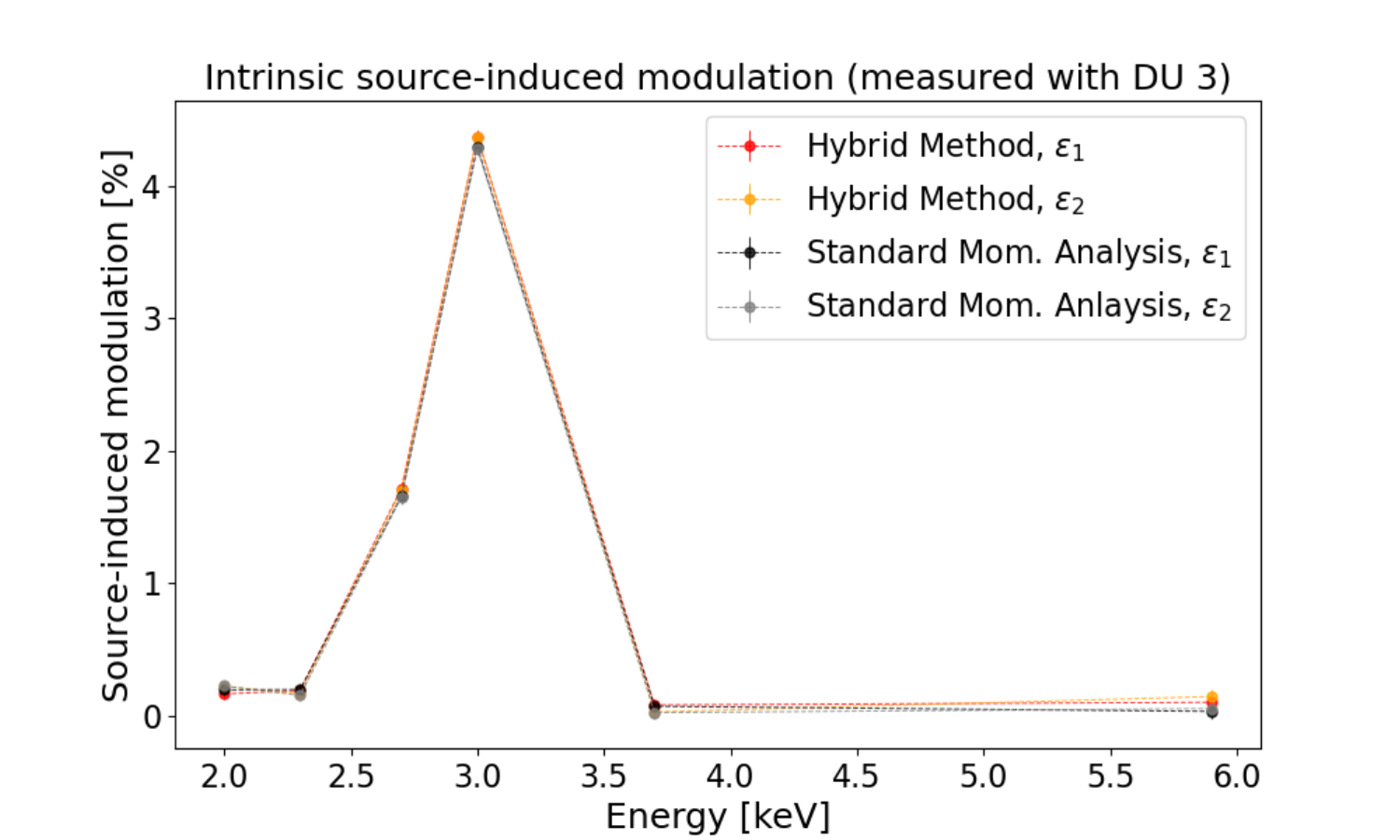} 
    \caption{Intrinsic source-induced modulation after the spurious modulation correction achieved with DU2 and DU3 data. Results are reported for each unpolarized energy beam, for both the hybrid method ($\epsilon_1$ configuration in \textit{red}, $\epsilon_2$ configuration in \textit{orange}) and the standard moment analysis ($\epsilon_1$ configuration in \textit{black}, $\epsilon_2$ configuration in \textit{gray}).}
    \label{fig:residual_m_du23}
\end{figure}

Fig.~\ref{fig:residual_m_du1} illustrates the residual modulation for DU1 after correcting with the spurious maps. The corresponding corrected results for DU2 and DU3 are shown in Fig.~\ref{fig:residual_m_du23}. Unlike the uncorrected results, these corrected values reflect only the intrinsic polarization of the sources, and we expect consistency across all DUs, which is indeed observed.

\section{High frequency peaks in the emission angles distribution}
\label{app:b}

\begin{figure}[htb]
    \centering
    \includegraphics[width=0.4\textwidth]
    {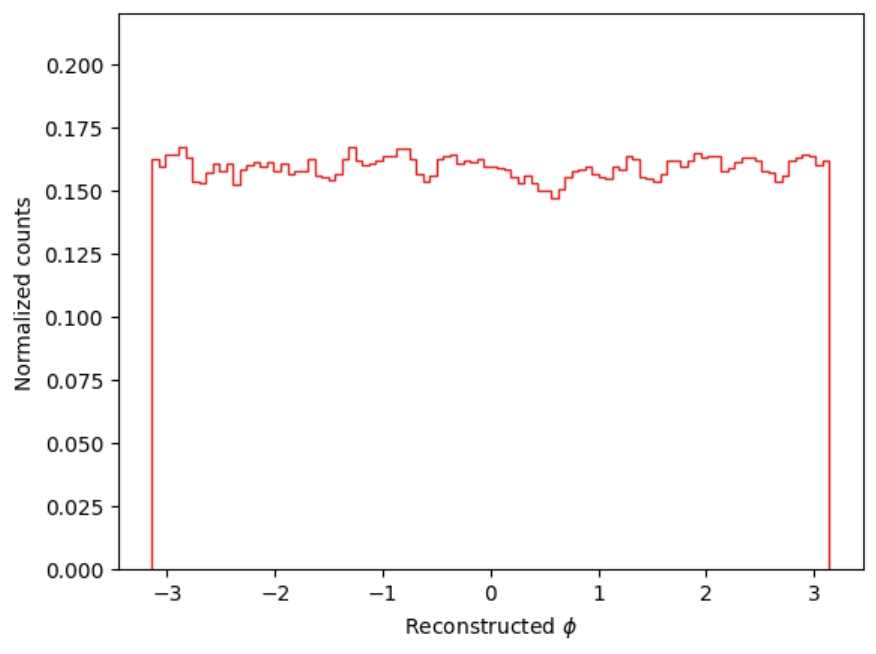}   
    \includegraphics[width=0.4\textwidth]
    {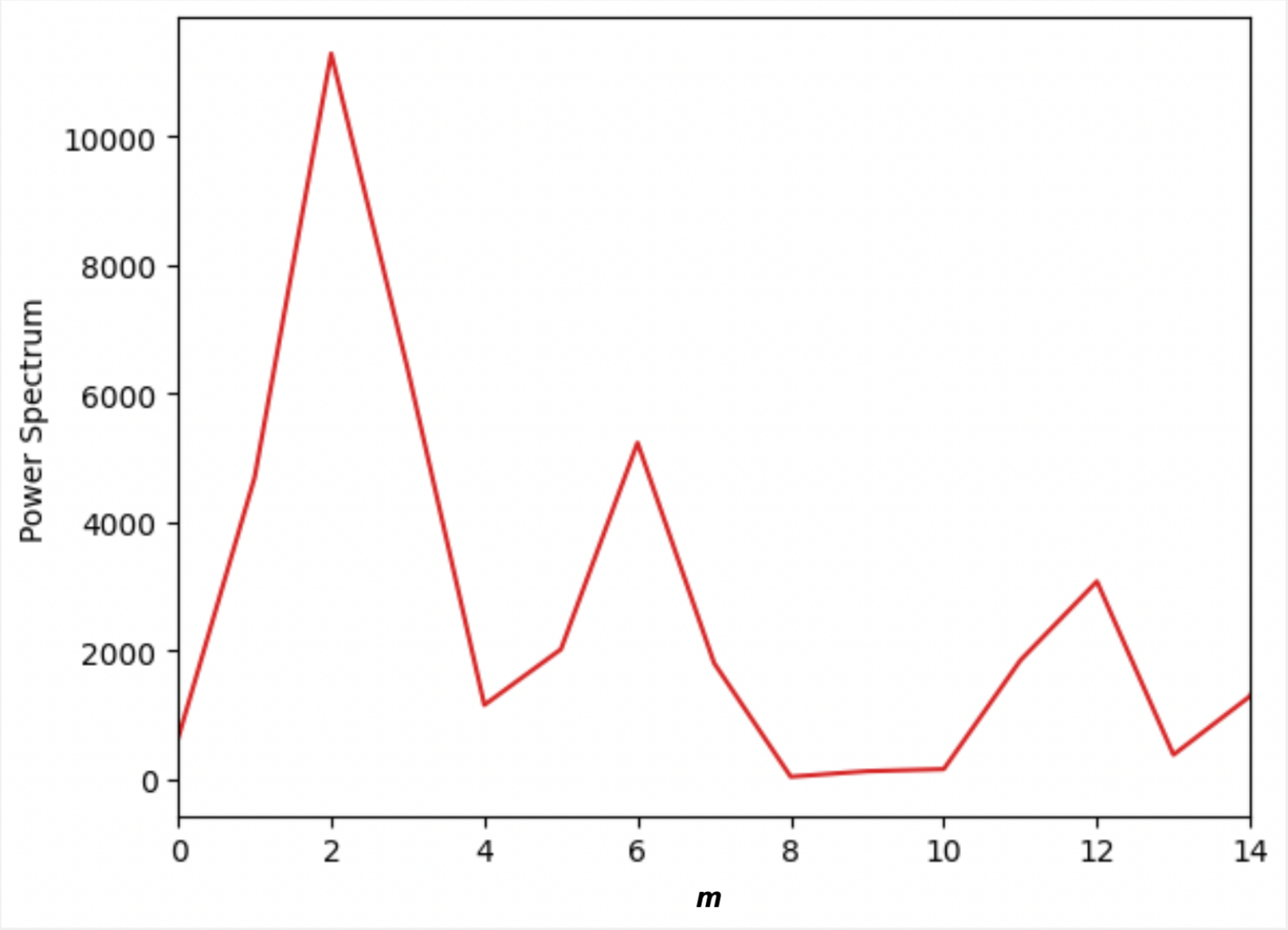}
    \caption{\textit{Left panel}: normalized distribution of the hybrid method predicted emission angles for a 2.7 keV beam subsample of $\epsilon_1$ configuration and DU1 data. \textit{Right panel}: power spectrum of the emission angles distribution.}
    \label{fig:power_spec_27}
\end{figure}

In Fig.~\ref{fig:mod_curves}, the distributions of the reconstructed emission angles for the three different unpolarized energy beams were reported for both the standard moment analysis in black and for the hybrid method in red. As already mentioned, excluding the mode $m$=2 peak due to the joint contribution of residual polarization and spurious modulation, higher frequency peaks ($m$=6,12) were detectable, especially for the hybrid method. 
For clarity, in Fig.~\ref{fig:power_spec_27} only the distribution of the hybrid method reconstructed emission angles for a 2.7 keV energy unpolarized beam (DU1 data, $\epsilon_1$ configuration), with the relative power spectrum, is reported. 

\begin{figure}[htb]
    \centering
    \includegraphics[width=0.4\textwidth]
    {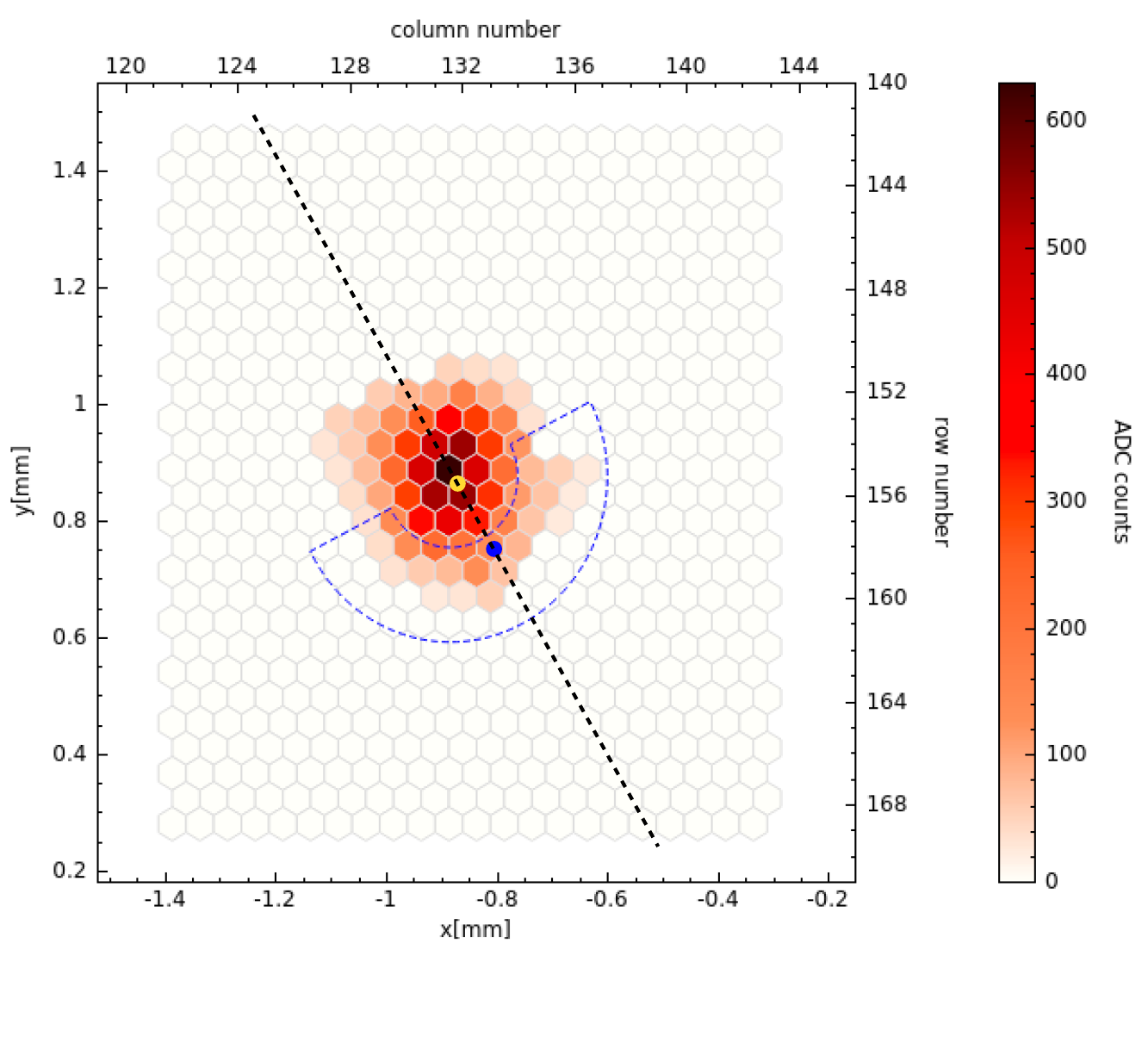}   
    \includegraphics[width=0.4\textwidth]
    {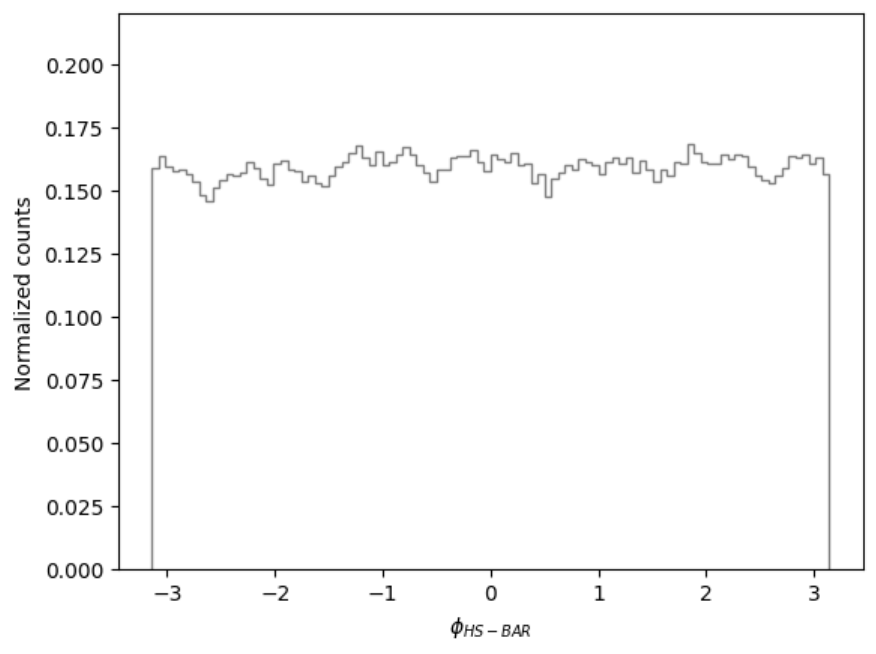}
    \includegraphics[width=0.4\textwidth]
    {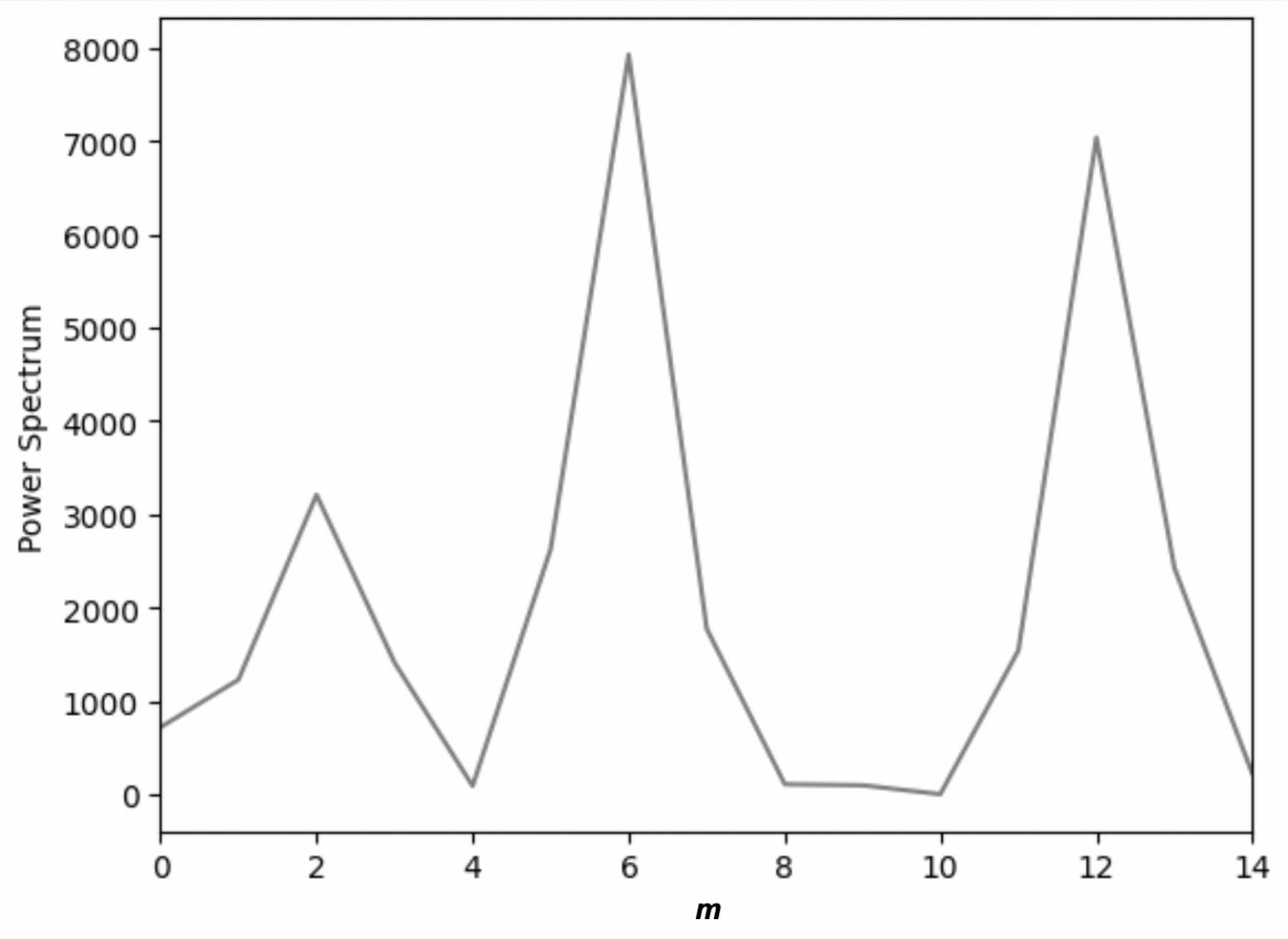}
    \caption{\textit{Upper left panel}: Example of a photoelectron track as measured by the GPD. The \textit{yellow dot} is the barycenter of the track, whereas the \textit{blue dot} is the barycenter of the horseshoe region (marked by the \textit{dashed blue line}). The \textit{dashed black line} is the direction of the vector connecting the two points. \textit{Upper right panel}: normalized distribution of the angles defined by the horseshoe barycenter (HS) and the track barycenter (BAR), denoted as $\phi_{HS-BAR}$, for the 2.7 keV energy beam. \textit{Lower panel}: power spectrum of the $\phi_{HS-BAR}$ distribution.}
    \label{fig:bary_hs}
\end{figure}

In order to understand the origin of the high frequency peaks, two features of the hybrid method need to be revisited.
Firstly, results regarding the impact point reconstruction showed the CNN inclination to predict the impact point close to the barycenter of the track for energies below 3 keV \citep{cibrario}. For low-energy tracks, in fact, the barycenter closely aligns with the true photon impact point.
Secondly, the final emission direction is assigned weighting the pixels charge according to their distance from the barycenter of the \textit{horseshoe region}, a region analytically evaluated to find the initial part of the track \citep{moment_analysis}, both in the standard moment analysis and in the hybrid method. The upper panel of Fig.~\ref{fig:bary_hs} reports a track example with its barycenter in yellow, whereas the blue dot is the barycenter of the horseshoe region (marked by a dashed blue line). The lower panels show the distribution of the angles defined by the vector connecting the horseshoe barycenter (HS) and the track barycenter (BAR) for the 2.7 keV energy beam (dashed black line in the upper panel), with its relative power spectrum. This distribution is entirely independent of CNN results and exhibits the same modes $m$ = 6,12 as the hybrid method emission angles distribution, and could explain the observed high frequencies peaks in the emission angle distribution.
Such peaks are less prominent in the standard moment analysis distribution, as the reconstructed impact point is generally further from the barycenter of the track. The origin of this biased alignment between the horseshoe barycenter and the track barycenter is linked to the hexagonal structure of the pixels, as both the mode $m$ = 6 and $m$ = 12 have phases which align with the symmetry axes of the GPD honeycomb matrix. However, this effect ultimately has no consequence on the determination of the Stokes parameters and, subsequently, on the polarization properties of the incident radiation.

\end{appendix}

\end{document}